\documentclass[iop]{emulateapj}
\pdfoutput=1
\usepackage{txfonts}
\usepackage{graphicx}
\usepackage{epstopdf}

\newcommand{\feh}{${\rm [Fe/H]}$}
\newcommand{\afe}{${\rm [\alpha/Fe]}$}
\newcommand{\mgfe}{${\rm [Mg/Fe]}$}
\newcommand{\sife}{${\rm [Si/Fe]}$}
\newcommand{\cafe}{${\rm [Ca/Fe]}$}
\newcommand{\tife}{${\rm [Ti/Fe]}$}

\shorttitle{Abundances from co-added Spectra}

\shortauthors{Yang et al.}

\begin{document}


\title{Measuring Detailed Chemical Abundances from co-added Medium
  Resolution Spectra. I. Tests Using Milky Way Dwarf Spheroidal
  Galaxies and Globular Clusters}

\author{Lei Yang\altaffilmark{1},  Evan N. Kirby\altaffilmark{2,3},  Puragra Guhathakurta\altaffilmark{4},  Eric W. Peng\altaffilmark{1,5},  Lucy Cheng\altaffilmark{6}}
%\author{Evan N. Kirby\altaffilmark{2,5}}
%\author{Puragra Guhathakurta\altaffilmark{3}}
%\author{Eric W. Peng\altaffilmark{1}}
%\author{Lucy Cheng\altaffilmark{4}}
%\author{Lei Yang\altaffilmark{1}}

\affil{\altaffilmark{1} Peking University, Department of Astronomy, 5 Yiheyuan Road, Haidian, Beijing 100871, China}
\email{yangleiseti@gmail.com}
\affil{\altaffilmark{2} California Institute of Technology, Department of Astronomy, Mail Stop 249-17, Pasadena, CA 91125, USA}
\affil{\altaffilmark{3} Hubble Fellow}
\affil{\altaffilmark{4} University of California Observatories/Lick Observatory, Department of Astronomy \& Astrophysics,
  University of California, 1156 High Street, Santa Cruz, CA 95064, USA}
\affil{\altaffilmark{5} Kavli Institute for Astronomy and Astrophysics, Peking University, 5 Yiheyuan Road, Haidian, Beijing 100871, China}
\affil{\altaffilmark{} }
\affil{\altaffilmark{6} The Harker School, 500 Saratoga Avenue, San
Jose, CA 95129, USA}

\begin{abstract}

The ability to measure metallicities and $\alpha$-element abundances in
individual red giant branch (RGB) stars using medium-resolution
spectra ($R\approx 6000$) is a valuable tool for deciphering the
nature of Milky Way dwarf satellites and the history of the Galactic
halo.  Extending such studies to more distant systems like
Andromeda is beyond the ability of the current generation
of telescopes, but by co-adding the spectra of similar stars, we can
attain the necessary signal-to-noise ratio to make detailed abundance
measurements.  In this paper, we present a method to determine
metallicities and $\alpha$-element abundances using the co-addition of
medium resolution spectra.  We test the method of spectral co-addition using
high-S/N spectra of more than 1300 RGB stars from Milky Way globular
clusters and dwarf spheroidal galaxies obtained with the Keck II telescope/DEIMOS
spectrograph \footnote{Data herein were obtained at the W.~M.~Keck
Observatory, which is operated as a scientific partnership among the
California Institute of Technology, the University of California, and
NASA. The Observatory was made possible by the generous financial
support of the W.~M.~Keck Foundation.}. We group similar stars using photometric criteria and
compare the weighted ensemble average abundances (\feh, \mgfe,
\sife, \cafe\ and \tife) of individual stars in each group with the measurements
made on the corresponding co-added spectrum. We find a high level of agreement between
the two methods, which permits us to apply this co-added spectra technique to
more distant RGB stars, like stars in the M31 satellite galaxies. This
paper outlines our spectral co-addition and abundance measurement methodology
and describes the potential biases in making these measurements.

\end{abstract}

\keywords{galaxies: abundances --- galaxies: dwarf --- galaxies: evolution --- galaxies: Local Group--- galaxies: stellar content}

\newpage

\section{Introduction}
		
The Local Group is dominated by the Milky Way and Andromeda galaxies,
with large families of dwarf satellite galaxies. The accessibility of
these fantastic targets, particularly the dwarf spheroidals (dSphs)
around the Milky Way, enables us to trace the dynamical and chemical
properties of individual stars in order to investigate
their formation. For decades, evidence has been growing that much of the material in
the Universe condensed into small dark matter halos at an early stage,
and over a Hubble time, many of these halos contribute to the growth
of massive galaxies in a ``chaotic accretion'' (Searle \& Zinn 1978;
White \& Rees 1978; Cole et al.\ 2000; Diemand et al.\ 2007). From observations, Helmi et al.\ (2006) showed
that there were metal-poor stars in the Milky Way halo with ${\rm [Fe/H]}< -3.0$~dex,
which did not seem to exist in dwarf spheroidal galaxies, and
therefore ruled out present day dSphs as the building blocks of the
Milky Way halo. Recently, however,
Kirby et al.\ (2008a, hereafter KGS08, 2009, 2010) determined the iron abundance distribution of vast majority of red giant branch
stars (RGBs) in globular clusters (GCs) and ultra-faint dSphs, and they found a significant metal-poor tail of stars does exist in
the Milky Way satellites, supporting the hierarchical formation of the stellar halo.

To investigate the formation of big spiral galaxies, especially their stellar halo accretion history,
chemical abundance patterns obtained from individual stars are crucial indicators
(Wheeler et al.\ 1989; Worthey 1994; Mannucci et al.\ 2010). Stars produce a diversity
of elements through nucleosynthesis, which are dispersed into the interstellar medium, and
are then mixed with material
in subsequent star formation. Generally, iron-peak elements, such as vanadium and iron, are
mainly generated by Type Ia supernovae (SNe Ia \textemdash Tinsley 1980) which are
important contributors to the total iron fraction in galaxies (Greggio
\& Renzini 1983).  Correspondingly, the $\alpha$-elements, like
oxygen, magnesium, silicon, calcium, titanium, etc., are produced in
core collapse supernovae (SNe II) whose progenitors are massive stars with typical stellar masses greater than
 $9 M_{\odot}$ (Wheeler et al.\ 1989; Woosley \& Weaver
1995; Gilmore 2004). Furthermore, compared to Type II supernovae, which have a
timescale of around $\sim 10$ million years (Pagel 1997; Woosley \& Janka 2005), Type Ia supernovae have a longer timescale of
at least $\sim 1$ Gyr (Matteucci \& Recchi 2001; Ishigaki et al.\ 2012). Hence, a plot of
\afe\ versus \feh\ tells us about the relative contribution of SNe Ia
and SNe II in star formation and evolution of a stellar system as a function of time, and
can be used as a clock to measure the intensity of star
formation at early stages. Additionally, comparing the distribution of
metallicity and $\alpha$-element abundances
between the different stellar components of the Milky Way and its companions
is a way of determining their evolutionary relationships.

Photometry is commonly used to determine the metallicities of old RGB
stars. This technique uses the locations of RGB stars in the color
magnitude diagram and compares them to empirical relations (Armandroff
et al.\ 1993), fitting functions (Saviane et al.\ 2000), or theoretical stellar tracks
(e.g., Harris \& Harris 2000; Mouhcine et al.\ 2005; Lianou et al.\ 2010) to derive
photometric estimates of the metallicity. The metallicities of distant stars in the Andromeda
system have also been estimated in this way.  Da Costa et al.\ (1996, 2000, 2002) used
HST/WFPC2 to measure photometric metallicities and obtained age estimates
using the RGB and horizontal branch in three dwarf spheroidal
galaxies, Andromeda I, Andromeda II, and Andromeda III, around M31. Kalirai et
al.\ (2010) studied hundreds of RGB stars in six M31 dSphs,
producing a luminosity-metallicity relation for dwarfs and compared it
with that of their Milky
Way counterparts. However, photometric metallicity estimates are often not
as accurate as spectroscopic metallicity for two main reasons. First, the $\alpha$
abundance of star often has to be assumed. Second, the
degeneracy of age and metallicity (Worthey 1994), even with resolved color-magnitude
diagrams, is difficult to untangle.
Lianou et al.\ (2011) compared photometric metallicity and spectroscopic
metallicity based on the near-infrared Ca triplet analysis of RGBs in five
Galactic dSphs, and they found that the agreement is good for
metallicities between $-2.0$~dex to $-1.5$~dex but that a high fraction of
intermediate-age stars would produce unreliable values.

Deriving chemical abundances using spectroscopy is much more reliable,
but also expensive in terms of data collection. A stellar spectrum contains a wealth of information,
allowing us to derive important stellar parameters like effective temperature,
surface gravity, metallicity and
other heavy elements abundances. Several empirical calibrations can be employed to measure
metallicity. Popular methods include \ion{Ca}{2}~K
$\lambda3933$ (Preston 1961; Zinn \& West 1984; Beers et al.\ 1999) and
the Ca near-infrared triplet calibration (Bica \& Alloin 1987;
Armandroff \& Zinn 1988; Olszewski et al.\ 1991; Rutledge et al.\ 1997; Foster et al.\ 2010) with
the prerequisite that \cafe\ must be assumed. Unfortunately, these calibrations fail when ${\rm [Fe/H]}< -2.2$
(Kirby et al.\ 2008b).

The most reliable way to measure abundances is to use high-resolution
spectroscopy ($R \gtrsim 20000$). This technique has been used to analyze
the detailed chemical abundance distribution of the Milky Way system and its
satellites (Shetrone et al.\ 1998, 2001, 2003; Venn et al.\ 2004; Tolstoy et
al.\ 2009; Letarte et al.\ 2010). High-resolution spectroscopic measurements of individual stars beyond
the Milky Way and its satellites, however, is extremely challenging.  Moreover,
high resolution spectroscopy is difficult to multiplex and often requires
observing one star at a time. For these reasons, medium-resolution
spectroscopy ($R\approx6000$), for which abundances of some elements
can still be well-measured, is an optimal choice for large sample
stars in Milky Way satellites, and even for the M31 system
(Guhathakurta et al.\ 2006; Koch et al.\ 2007).

Previous studies have targeted stars in the Milky Way satellites
(e.g., Lanfranchi \& Matteucci 2004;
Shetrone et al.\ 2009; Strigari et al.\ 2010). The largest sample was given by KGS08 and Kirby
et~al.\ (2009, 2010) who observed more than
2500 RGB stars in Milky Way GCs and dSphs at medium-resolution
using Keck/DEIMOS (Faber et al.\ 2003). They performed
a multi-element abundance analysis using
pixel-to-pixel matching based on a grid of synthetic stellar spectra.
They verified their abundance measurement
technique by comparing their results with the abundances
derived from high-resolution spectra. In their study, in addition to finding a long metal-poor tail in
ultra-faint dSphs, matching that of the Galactic halo, they derived
trends for individual $\alpha$-elements (Mg, Si, Ca, Ti) versus
metallicity, demonstrating
that there was not much metal enrichment before the onset of SNe~I
(Kirby et al.\ 2011a, 2011b).

Observations of M31's halo present a more
complex accretion history (Ibata et al.\ 2001; Choi
et al.\ 2002; Reitzel \& Guhathakurta 2002; McConnachie et al.\ 2004;
Guhathakurta et al.\ 2006, 2010;  Kalirai et al.\ 2009, 2010;
Collins et al.\ 2010; Tanaka et al.\ 2010), indicating that a study of abundance trends in its satellites could be more interesting
and challenging. However, the larger distance and fainter
apparent magnitudes hamper any detailed
investigation of individual stellar abundances for dwarf satellites in
M31, although there have been attempts to obtain detailed chemical abundance patterns in M31 globular clusters using high-resolution,
integrated-light spectroscopy (Colucci et al.\ 2009).
One way to address this problem
is by co-adding many similar stars to produce spectra with higher
signal-to-noise (S/N).  This process, if done properly, would extend
our ability to obtain elemental abundance analysis to larger distances, but there is also
the potential for introducing biases.

The co-addition of spectra to measure the ensemble properties of
similar objects has a long history in spectral
processing.  Adelman \& Leckrone (1985) used co-addition for the ultraviolet
and optical region of spectrum. Holberg et al.\ (2003) co-added
multiple observations of individual white dwarf stars to enhance the
signal-to-noise ratio and combined them into a single spectrum. Gallazzi
et al.\ (2008) used co-added spectra of galaxies with similar velocity
dispersions, absolute r-band magnitude and 4000$\text{\AA}$-break values for
those regions of parameter space where individual spectra had lower S/N.
Most recently, Schlaufman et al.\ (2011, 2012) compared the
average metallicities and $\alpha$-element abundances between the
elements of cold halo substructure (ECHOS) and the kinematically smooth
stellar inner halo along lines of sight in the Sloan Extension for Galactic
Understanding and Exploration (SEGUE) by co-adding spectra
of the metal-poor main sequence turnoff stars identified by the SEGUE Stellar
Parameter Pipeline. Using noise-degraded spectra, they found that the
mean square error (MSE) of abundances derived from co-added
spectra is from 0.05 dex for metal-rich (iron-rich) stars
to 0.2 dex for \feh\ and \afe\ for the most metal-poor stars. In a
test using the globular clusters M13 and M15,
the MSE roughly equals to 0.1 dex for most of the metal-rich stars
and to 0.2 dex for metal-poor stars, in both \feh\ and
\afe. However, the individual abundances of $\alpha$-elements
are inaccessible in their measurements. In this work, we use weighted
spectral co-addition of RGB stars that share similar intrinsic
properties to increase the S/N so that
individual $\alpha$-elements can be measured.

All data used in this paper are from the same data sets used in KGS08
and Kirby et al.\ (2009, 2010),
which involves thousands RGB stars in Milky Way GCs and dSphs. We test
our co-addition method by measuring metallicity (\feh) and four
$\alpha$-element abundances (\mgfe, \sife, \cafe, \tife). We used the same definition of chemical abundances as
defined in KGS08 \footnote{In their work,
$\rm 12+log[n(Fe)/n(H)]_{\odot}=7.52$ is adopted in \S\,4 of KGS08, where n is
number density (as adopted by Sneden et al.\ 1992). The abundances of other elements adopted solar composition
from Anders \& Grevesse (1989) in Table 8 of Kirby et al.\ (2010)}.

First, we present a short summary of the observations and data
reduction in \S2. In \S3 we describe the method of abundance
determination using a synthetic spectral grid, as well as our method
of co-addition: star selection and grouping, weighted co-adding, and
abundance measurement. In \S4, we present the
comparison of weighted ensemble average of individual abundances to
abundances measured from
co-added spectra. We also discuss discrepancies and biases between
the two results. We summarize our work in \S5.

\section{Observation and Data Reduction}
All the medium-resolution spectra of RGB stars used in this study are from
KGS08 and Kirby et al.\ (2009, 2010),
which include 2947 RGB stars in 8 Milky Way dSph galaxies
and 654 RGB stars in 14  Galactic
globular clusters. The observations were performed with DEIMOS on the
Keck~II telescope. The spectrograph configuration used the OG550 filter with the 1200
line $mm^{-1}$ grating at a central wavelength of $\sim\!\!7800\text{\AA}$
with a slit width of $0\farcs7$. The spectral resolution is
$\sim\!\!1.2\text{\AA}$ to $\sim\!\!1.3\text{\AA}$
(corresponding to a resolving power $6500<R<7000$ at $8500\text{\AA}$)
with a spectral range of 6300--9100$\text{\AA}$. Exposures of Kr, Ne, Ar,
and Xe arc lamps were used for wavelength calibration and exposure of a quartz
lamp provided the flat field. The DEIMOS data reduction pipeline
developed by the DEEP galaxy redshift
survey\footnote{http://astro.berkeley.edu/~cooper/deep/spec2d/  \\
The analysis pipeline used to reduce the DEIMOS data was developed at UC Berkeley with support from NSF grant AST-0071048.} was used to extract one
dimensional spectra (Newman et al.\ 2012; Cooper et al.\ 2012). The pipeline traced the edges of slits in the flat
field to determine the CCD location of each slit. A polynomial fit to
the CCD pixel locations of arc lamp lines provided the wavelength
solution. Each exposure of stellar targets was rectified and then sky
subtracted based on a B-spline model of the night sky emission lines.
Then, the exposures were combined with cosmic ray rejection into one
two-dimensional spectrum for each slit. Finally, the one-dimensional
stellar spectrum was extracted from a small spatial window encompassing
the light of star in the two-dimensional spectrum. The product of the
pipeline was a wavelength calibrated, sky-subtracted,
cosmic-ray-cleaned, one dimensional spectrum for each target. A hot star
template spectrum was employed to remove the terrestrial atmospheric
absorption introduced into the stellar spectra. In continuum
determination, a B-spline was used to fit the ``continuum
regions"\footnote{They call spectral regions with synthetic flux greater
than 0.96 and a minimum width of 0.5 $\text{\AA}$ ``continuum regions''.} of
the spectra. Each pixel was weighted by its inverse variance in the
fit, and the  fit was performed iteratively such that pixels that deviated
from the fit by more than 5$\sigma$ were removed from the next iteration
of the fit. For further details, please see KGS08.  Table~\ref{tab:rgbsource}
lists all the stellar systems used in this study.

\begin{table}[h]\label{tab:rgbsource}
\begin{center}
\centering
\caption{RGB spectral data sets}
\label{tab:rgbsource}
\centering
{\small
\begin{tabular}{lcc}
\hline \hline
  Name   &   \multicolumn{2}{c}{Number of RGB stars}\\
             &    Total & $\log g \leq 1.4$ \\
\hline
 Globular Clusters & \\
\hline
  NGC 6205 (M13)            &  68   &  9   \\
		NGC 7078 (M15)          &  134  &  26  \\
		NGC 1904 (M79)          &  58   &  17  \\
		NGC 2419               &  95   &  40  \\
		NGC 7006               &  11   &  9   \\
		NGC 7492               &  20   &  5   \\
		NGC 5024 (M53)\tablenotemark{$\bigstar$}          &  49   &  2   \\
		NGC 6656 (M22)\tablenotemark{$\bigstar$}          &  50   &  1   \\
		NGC 5053\tablenotemark{$\bigstar$}               &  49   &  1   \\
		NGC 288\tablenotemark{$\bigstar$}                &  30   &  0   \\
  NGC 5904 (M5)           &  51   &  6   \\
		NGC 6838 (M71)\tablenotemark{$\bigstar$}          &  34   &  1   \\
		NGC 7089 (M2)           &  44   &  5   \\
		Pal 13\tablenotemark{$\bigstar$}              &  10   &  0   \\

\hline
  dSphs & \\
	\hline
		Canes Venatici I &      174  &  36  \\
		Draco            &      298  &  38  \\
		Fornax          &       675  &  280 \\
		Leo I           &       813  &  571 \\
		Leo II          &       258  &  119 \\
		Sculptor        &       376  &  153 \\
		Sextans         &       141  &  16  \\
		Ursa Minor      &       212  &  21  \\
\hline \hline
\tablenote{$\bigstar$ These globular clusters were not used in this work because there were not enough RGB stars for coaddition.}
%\begin{tablenotes}
%\footnotesize
%\item $\bigstar$These globular clusters were not used in this work because there were not enough RGB stars for coaddition.
%\end{tablenotes}
\end{tabular}
}
%\caption{$\bigstar$ These globular clusters were not used in this work because there were not enough RGB stars for coaddition.}
%\begin{tablenotes}
%\footnotesize
%\item Globular clusters which are noted by $\ast$ are not used in this
%work for there is not enough RGB stars for co-addition.
%\end{tablenotes}
\end{center}
\end{table}
\section{Abundance Measurements}

\begin{table*}[t]
\begin{center}
\centering
\caption{Atmospheric Parameter Grid.}
\label{tab:grid}
\centering
{\small
\begin{tabular}{lcccclllll}
\hline \hline
  Quantity                  &  Minimum  &  Maximum  &  Step                        &  Number \\
\hline
$\it T_{\rm eff}(\rm K)$              &  3500   &  8000    &  100 ($\it T_{\rm eff} \leq \rm 5600$)      &22 \\
\quad                     &         &          &  200 ($\it T_{\rm eff} \ge \rm 5600$)     & 12 \\
 $\log \it g$ (cm $s^{-2}$), $\it T_{\rm eff} \le \rm 6800$ & 0.0 & 5.0 & 0.5                  & 11 \\
 $\log \it g$ (cm $s^{-2}$), $\it T_{\rm eff} \ge \rm 7000$ & 0.5 & 5.0 & 0.5                  & 10 \\
\ [M/H] (atmosphere)  & $-4.0$    & 0.0          & 0.5                          & 9  \\
\ [M/H] (spectra)      & $-4.0$    & 0.0          & 0.1                          & 41 \\
\ \afe       & $-0.8$    & $ +1.2 $         & 0.1                          & 21 \\
\hline \hline
\end{tabular}
}
\end{center}
\end{table*}

%%%%%%%%%%%%%%%%%%%
%%%%%%%%  Figure 1
%%%%%%%%%%%%%%%%%%%
\begin{figure*}

  \plotone{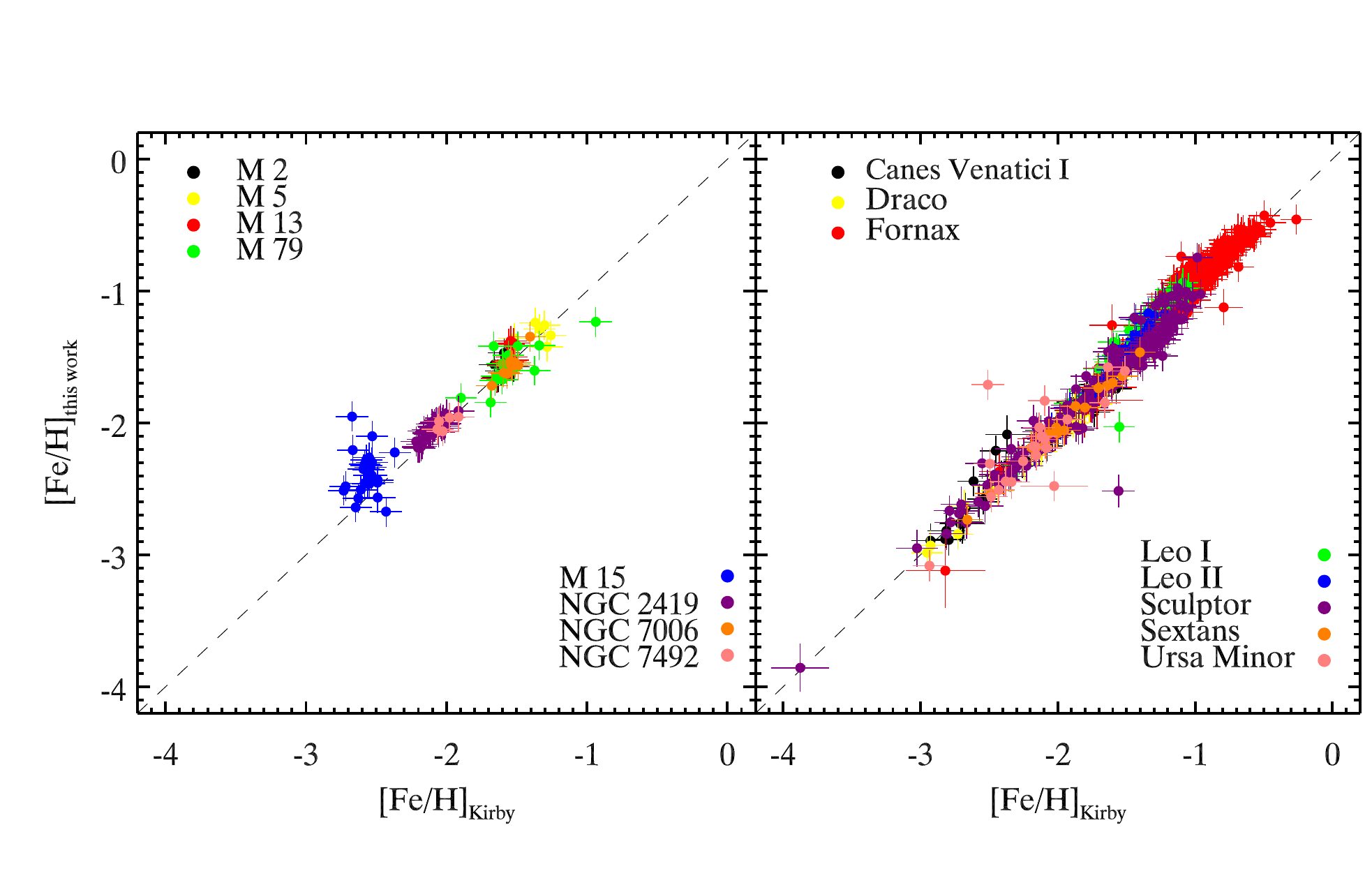}
  \caption{The comparison of metallicity measured from medium-resolution
    spectra for individual stars between
		Kirby et al.\ and our codes, developed for co-added
		spectra. The left panel presents results for 8 GCs and the right panel
		for 8 dSphs. All these RGB stars meet the $\log \it g \rm \leq 1.4$ threshold.
		The key difference between these two measurements for individual stars
		is in our codes we used a fixed $\it T_{\rm eff}$ and $\log g$ derived from
		photometric estimates versus allowing $\it T_{\rm eff}$ to
		float during the spectral fitting in Kirby et al.'s codes. The
		Levenberg-Marquardt algorithm is adopted to find the best-fitting
		spectrum with several iterative steps then give the abundance. The
		error bars shown include both the random errors in the fit and the
		systematic errors adopted from Kirby et al..
		\label{fig:comparison}}
\end{figure*}

Kirby et al.\ presented a technique for multi-element
abundance measurements of medium-resolution spectra, which enable them to
determine individual $\alpha$-element abundances of RGB stars in
the Milky Way globular clusters and dwarf
satellite galaxies. In this technique a large grid of synthetic spectra is used, so there is no restriction imposed on the
metallicity range, overcoming the problems encountered by other methods. In brief,
the photometric effective temperature ($\it T_{\rm eff}$) and surface
gravity ($\log \it g$) are determined from isochrone-fitting on the
color-magnitude diagram using three different model isochrones (Kirby
et al.\ 2009)---Yonsei-Yale (Demarque
et al.\ 2004), Victoria-Regina (VandenBerg et al.\ 2006),
and Padova (Girardi et al.\ 2002)---and an empirical color-based $\it T_{\rm
eff}$ (Ram\'{i}rez \& Mel\'{e}ndez 2005).
Then, they adopted the Levernberg-Marquardt
algorithm ({\tt mpfit}, written by Markwardt 2009)
to find the best-fitting synthetic spectrum to the observed
spectrum in several iterative steps by minimizing the $\chi^{2}$
calculated from the degraded synthetic spectrum and the observed
spectrum. Lastly, the stellar parameters of the best-fitting synthetic
spectrum is presented as the observed one's. With different elemental masks, the abundances measured
include iron abundance (\feh) and four $\alpha$ elements (Mg,
Si, Ca and Ti).
%The synthetic spectra had only one dimension of $\alpha$ abundances.
The four individual $\alpha$ elements were measured by
considering only spectral regions most sensitive to the corresponding element.  For
example, the Mg wavelength mask covers about 20 neutral Mg lines in
the DEIMOS spectral range.  Even though all of the $\alpha$ elements vary
together in the synthetic spectrum, only Mg lines are used to
determine \mgfe. For those are interested in elemental masks, KGS08\footnote{Spectral mask, TABLE 2} and
Kirby et al.\ (2009)\footnote{Section 4.6 and Figure 5} depicted more details on the
construction of the wavelength masks. They performed extensive comparisons of their medium resolution
results with high-resolution spectroscopic elemental
abundances from previous studies to validate their technique.
In this study, we inherit the idea of Kirby's technique and make proper modification
to meet our demands to determine chemical abundances of more distant RGB stars beyond the Milky Way.
But firstly, we aim to test our co-add spectra method with the medium-resolution spectra used by Kirby et al.,
and the membership of these RGB stars has been confirmed by Kirby et al..

\subsection{Synthetic Spectral Library}

We accomplished all chemical abundance analysis on the basis of a large grid
of synthetic spectra generated by Kirby et al.\ (KGS08, 2009, 2010). Based on ATLAS9 model atmospheres (Kurucz
1993; Sbordone et al.\ 2004; Sbordone 2005) without
convective overshooting (Castelli et al.\ 1997, 2004; Castelli 2005),
and a line list of
atomic and molecular transition data from  the Vienna Atomic Line
Database (Kupka et al.\ 1999), Kirby et al.\ synthesized
spectra using the local equilibrium, plane-parallel spectrum synthesis
code MOOG (Sneden 1973), which span the same wavelength range as the data (6300
to 9100$\text{\AA}$) with a resolution of 0.02$\text{\AA}$. The synthetic spectra
fail in modeling of the \ion{Ca}{2} triplet, \ion{Mg}{1}
$\lambda$8807 and the absorption lines of TiO. To avoid an unexpected
discontinuity they recomputed new ODFs for the new grid with DFSYNTHE
code (Castelli 2005) and employed the solar composition of
Anders \& Grevesse (1989), but for Fe they used Sneden et al.\ (1992)
(Kirby et al.\ 2009).
To prevent unwittingly discarding the extremely metal-poor stars
beyond the preliminary boundary of grid, they also expanded the
synthetic spectral grid limit to $\feh =-5.0$ (Kirby et al.\ 2010). The value of
\afe\ for the stellar model atmospheres
would be different for each individual $\alpha$-elements because elements
have been measured only with the spectral regions where are the most sensitive
to the corresponding element. Therefore, an
additional subgrid with the extra dimension of $\alpha$-element
abundance (\afe$_{\rm abund}$) is also generated for more
accurate measurement of
\mgfe, \sife, \cafe\ and \tife\ at fixed \afe$_{\rm atm}$. This spectral library includes four dimensions:
effective temperature ($\it T_{\rm eff}$), surface gravity ($\log \it g$),
metallicity (\feh\ ), $\alpha$ abundance (\afe$_{\rm atm}$) of the stellar
 atmosphere. Table~\ref{tab:grid} gives the limited ranges and steps of these five parameters.
This spectral grid is available online, and readers with
interests in this grid are recommended to refer to more details in Kirby (2011c).
%~\ref{tab:grid} 

\subsection{Individual Stellar Abundances}
\label{sec:indabund}

To determine stellar abundances, we developed an independent code
based on Kirby et al.'s technique with some refinements for the application
to co-added spectra. For testing purposes,
most of the medium-resolution spectra used have $\rm S/N>20/pixel$ with mean
S/N around 80/pixel.

In spectral co-addition, an essential step is to rebin the spectra in preparation for
co-adding. Our approach rebins each science
spectrum onto a common wavelength region (6300--9100$\text{\AA}$ with step
0.25$\text{\AA}$). The same has been done to the degraded synthetic
spectrum which is going to be compared with the rebinned science one.
Considering our co-added spectra approach aims to be applied on the RGB stars
 of M31 dwarf satellite galaxies whose spectroscopic temperatures are
not available, we fixed the effective temperature, as well
as the surface gravity, with the value derived from photometry. When we measured the
effective temperature and surface gravity, the Yonsei-Yale isochrones
fitting was carried out on the CMD at an assumed age of 14 Gyr and \afe\ $= + 0.3$
for all RGB stars, however, in Kirby et al.'s work they only estimated $\log \it g$ by photometry.
After setting the initial parameters, we performed the abundance determination on the rebinned spectra.
In order to verify that our method works well on individual
stars, we redetermined chemical abundances of all RGB stars with our code.
 Figure~\ref{fig:comparison} shows the comparison of \feh\ between Kirby's and ours.
The stars in the Figure~1 are also used for later co-addition test,
but some stars whose spectra had insufficient S/N to measure a particular element have been removed.
The selection detailed is discussed in next section \S3.3.
%~\ref{fig:comparison} 

\subsection{Surface Gravity Restriction}
\begin{figure}

  \plotone{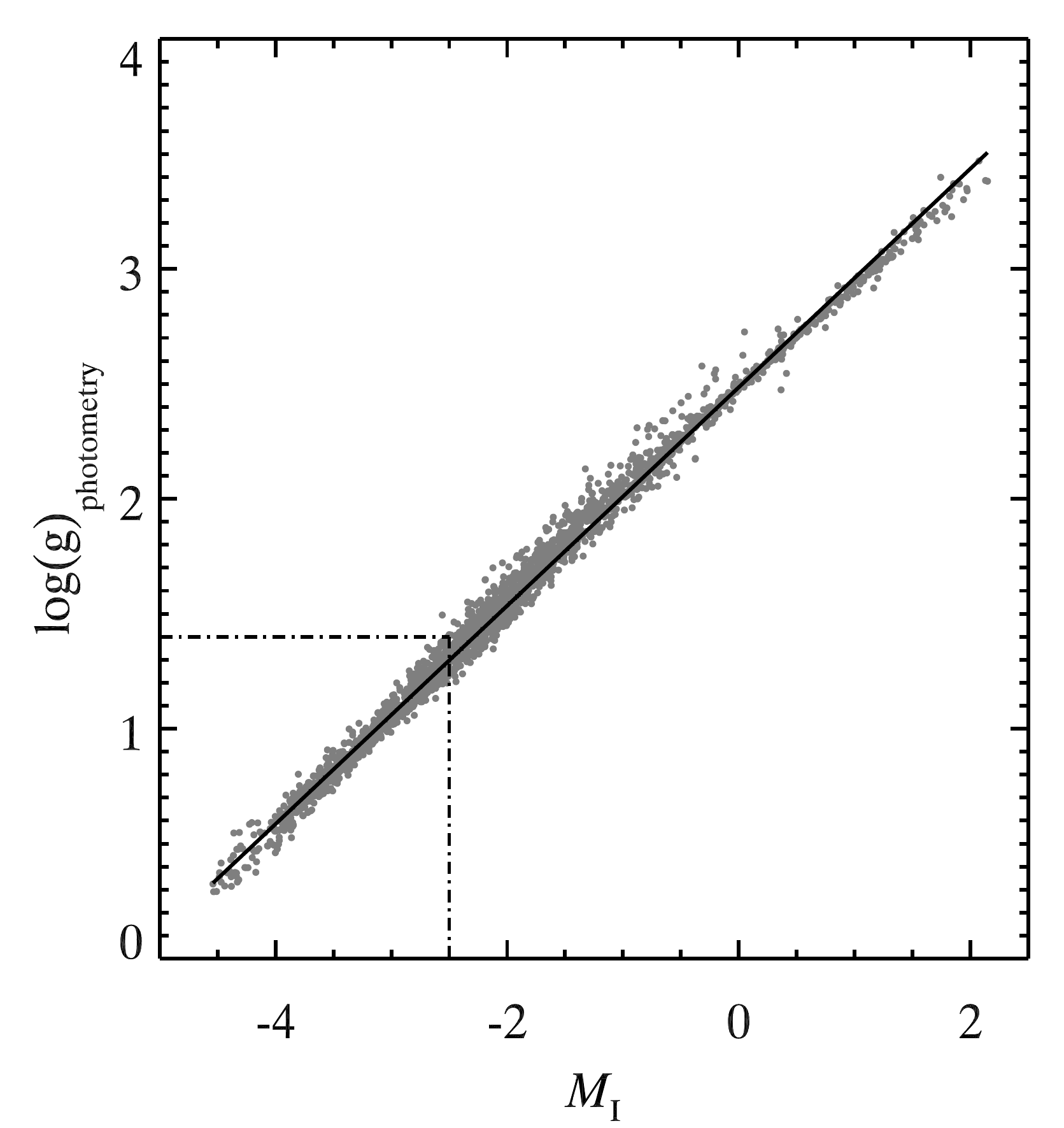}
  \caption{We use a linear relationship between photometric surface
		gravity ($\log \it g$) and $I$-band absolute magnitude
                ($\it M_{\rm I}$), which is converted from
		apparent magnitude with the extinction-corrected distance modulus, to
		select stars whose $\log \it g$ $\rm \leq 1.40$ ( $(\it M_{\rm I})$ $\geq
		-2.5$ ). The plot contains all dSph RGB stars except for
		Fornax for which we have no $I$-band data. The line is
                the best fit linear relation.
		Stars with $\log \it g\rm \leq 1.4$ are used to produce co-added
                spectra.
                \label{fig:milogg}}
\end{figure}

To start, we performed chemical abundance determination for more
than 3600 RGB stars from 14 globular clusters and 8 dwarf spheroidal galaxies.
The stellar ages of these RGB stars are difficult to estimate but fortunately
have only a small impact on the measured chemical abundances (Harris et
al. 1999; Frayn \& Gilmore 2002; Lianou et al. 2010), so we
assumed an age of 14 Gyr for all RGB stars (Grebel \& Gallagher 2004)
and set \afe\ $ = + 0.3 $ empirically.
Then, the effective temperature $(\it T_{\rm eff})$ and
surface gravity ($\log \it g$) of member stars were estimated by fitting Yonsei-Yale
isochrones on the color-magnitude diagram. We then proceeded to measure the
individual abundances as described in \S3.2. We found some element
abundances of some stars were unmeasurable, and we expected that low S/N is a
possible reason. Thus, for the purposes of this test, stars for which
we cannot measure a particular element abundance were not used in the
co-addition. Additionally, previous observations of RGB stars in M31 showed that only stars with
$\it M_{\rm I} \rm \leq -2.5$ are accessible for spectroscopy. So we further introduced a
cut in $\log \it g$. Given the roughly linear
relationship between $\it M_{\rm I}$ and $\it T_{\rm eff}$, this corresponds to a
selection in $\log \it g$.  Figure~2 shows the
linear relationship between photometric $\log \it g$ and absolute magnitude
in the $I$-band ($\it M_{\rm I}$) for 7 dSphs (except Fornax, for which we
do not have $I$-band data).
 From Figure~\ref{fig:milogg} $\it M_{\rm I} \rm =-2.5$ 
 roughly corresponds to a cut at $\log \it g \rm = 1.40$. Thus, we only selected stars 
having photometric $\log \it g$ $\rm \leq 1.40$ for the co-addition. The
number of stars left for each dSph and GC after
this selection is listed in Table~\ref{tab:nstars}. Only 8 globular
clusters have enough stars for the following test.
%~\ref{fig:milogg}
%~\ref{tab:nstars}

%%%%%%%%%%%%%%%%%%%%%%
%%%%%%%%% Figure 2
%%%%%%%%%%%%%%%%%%%%%%

%%%%%%%%%%%%%%%%%%%%%%%%%%%%%%%%%%%%%%%%%%%%%
\begin{table}[h]
\begin{center}
\centering
\caption{Star groups}
\label{tab:nstars}
\centering
{\small
\begin{tabular}{llllllllll}
\hline \hline
  Name of Galaxies  &  $N_{bin}$\tablenotemark{a} &	$N_{good stars}$\tablenotemark{b}   \\

\hline
 Globular Clusters & \\
\hline
  NGC6205(M13)  &   1      &  9   \\
		NGC7078(M15)  &   4      &  5  \\
		NGC1904(M79)  &   3      &  5  \\
		NGC2419       &   5      &  5  \\
		NGC7006       &   1      &  9  \\
		NGC7492       &   1      &  5   \\
  NGC5904(M5)   &   1      &  6   \\
 	NGC7089(M2)   &   1      &  5   \\

\hline
  dSphs & \\
	\hline
		Canes Venatici I & 4     &  5  \\
		Draco            & 6     &  5  \\
		Fornax          &  30    & 8 \\
		Leo I           &  44    & 10  \\
		Leo II          &  11    & 8 \\
		Sculptor        &  16    & 8 \\
		Sextans         &  2     &  5  \\
		Ursa Minor      &  2     &  5  \\
\hline
\tablenote{$N_{bin}$ is the number of bins for each stellar system.}
\tablenote{$N_{good stars}$ is the minimum number of good stars we set for each bin.}
\end{tabular}
}
%\caption{$~^aN_{bin}$ is the number of bins for each stellar system.}
%\caption{$~^bN_{good stars}$ is the minimum number of good stars we set for each bin.}
\end{center}
\end{table}
%%%%%%%%%%%%%%%%%%%%%%%%%%%%%%%%%%%%%

\subsection{Grouping and co-addition}
\label{sec:co-add}

We consider photometric effective temperature ($\it T_{\rm eff}$) and the
photometric metallicity estimate ($\rm [Fe/H]_{\rm phot}$), respectively, to organize the remaining
stars into groups for co-adding. The photometric metallicities are also
derived by Yonsei-Yale theoretical isochrones fitting with an age of 14
Gyr (See \S3.3). We used a cut at
$\log \it g$ $\rm \leq 1.40$ to ensure that all stars lie in a range of about 1
dex in $\log \it g$. Moreover, the synthetic spectral measurements use
neutral metal lines only which are nearly insensitive to surface gravity.
Therefore, $\log \it g$ barely changes the strength of spectral features,
making it acceptable  not to include $\log \it g$ in the binning.

The goal of this study is, for each grouping, to compare the weighted
average abundances of the individual stars (the input) with the
abundances measured on the co-added spectra (the output).  For this
purpose, it is important that each star used has a measurable
abundance.  Our pipeline is able to measure \feh\ in all sample stars, but for some of them,
individual $\alpha$ elements (e.g.,
Mg) may not be measurable due to the quality of the spectra.
Including these spectra in the grouping would bias the measurement on
the co-added spectrum (but contribute nothing to the weighted average),
so we are very careful to construct separate groupings for each
element measured, i.e., a co-added spectrum for testing \mgfe\ consists
only of stars that have reliable \mgfe\ measurements individually. This allows us to use the maximum number of available
stars for testing each element.

\begin{figure}

  \plotone{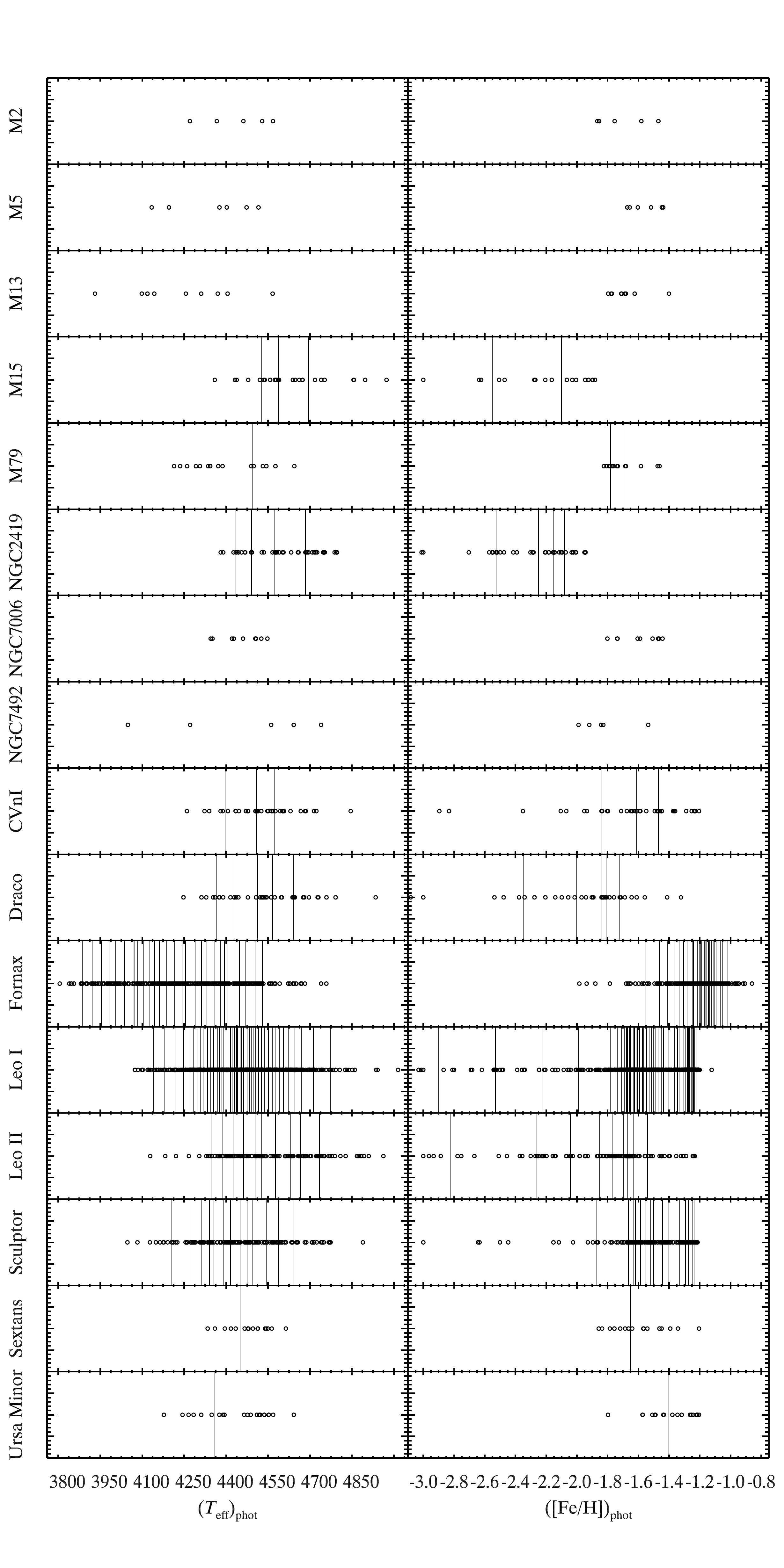}
  \caption{ Two scenarios have been used to bin stars. The open circles are individual stars used in this test.
            The vertical lines split the test stars into groups by two ways. Left panels show stars binned
            by photometric effective temperature. Right panels show stars binned by photometric metallicity. There
            are five GCs (M2, M5, M13, NGC7006, and NGC7492) have one bin for the limited number of stars. From this
            comparison, the dSphs, which are most affected by the two scenarios, are Fornax, Leo I,
            and Sculptor. We detail the impact of different binning schemes
            in the discussion section.  \label{fig:binplots}}
\end{figure}

After ranking member stars by their $(\it T_{\rm eff})_{\rm phot}$ and
$\rm [Fe/H]_{\rm phot}$ separately, we make
sure that each group contains at least 5 stars for which all five elemental
abundances are measurable individually. We set 8 as the minimum number
of stars for Fornax, Leo II and Sculptor, and 10 for Leo I for their large number
of stars. Table 3 lists details for each GC and
dSph. When we co-add spectra together for one bin, we produce five
different co-adds, that is for each elemental abundance
of \feh, \mgfe, \sife, \cafe, and \tife, we only add the
spectra whose elemental abundance is measurable individually, and use
that co-added spectrum to determine the corresponding elemental
abundance. The bad spectral regions therefore make no contribution to
the co-added spectrum for the element of interest, which is
equivalent to the elemental abundance derived from the weighted
ensemble average. Figure~\ref{fig:binplots} shows two binning scenarios in detail for 8 GCs and 8 dSphs,
by photometric effective temperature and metallicity. As the left panels show, stars having expanded distribution in
 $\it T_{\rm eff}$ are binned more evenly, especially for Fornax, Leo I, and Sculptor, whose
$\rm [Fe/H]_{\rm phot}$ are more concentrated. The impacts of this difference to the measured metallicities
for these bins are clearly shown in the later figures.

%%%%%%%%%%%%%%%%%%%%%%
%%%%%%%%% Figure binning
%%%%%%%%%%%%%%%%%%%%%%

%%%%%%%%%%%%%%%%%%%%%%%%%%%%%%%%%%%%%%%%%%%%%

For the observed spectrum, we use pixel masks to remove bad
spectral regions, like telluric absorption and cosmic rays, before rebinning.
Keck/DEIMOS has eight CCD and the whole spectrum spans two CCDs, so we exclude 5 pixels near the end of each
CCD which may cause artifacts. We then rebin the spectrum
onto a common wavelength range (6300$\text{\AA}$ to 9100$\text{\AA}$) and add
fluxes of the normalized rebinned spectra together weighted by the rebinned inverse variance.
The co-addition equation is:

\begin{equation} \label{eq:co-add}
{\overline x_{pixel,i} =\frac{\sum_{j=1}^{n}{(x_{pixel,ij}/\sigma_{pixel,ij}^{2})}}{\sum_{j=1}^{n}{1/{\sigma_{pixel,ij}^{2}}}}}
\end{equation}

where $x_{pixel,ij}$ represents the flux in the $i$-th pixel of $j$-th spectrum in a group of stars,
$\sigma_{pixel,ij}^2$ is the variance of $x_{pixel,ij}$, $n$ represents the total
number of spectra in the group, $\overline x_{pixel,i}$ is the weighted average flux of $i$-th pixel of $n$ spectra.

For consistency, we also
create a grid of co-added synthetic spectra. %for abundance measurements of the co-added science spectra.
First, for one group of stars, we pick the same number of synthetic spectra
with same chemical abundances but with different $\it T_{\rm eff}$ and $\log \it g $.
 The synthetic spectra are chosen to have the $\it T_{\rm eff}$ and
$\log \it g $  as determined by the photometric estimates of
 the observed stars. Second, we smoothed all the synthetic spectra with a
Gaussian filter to match the
spectral resolution science spectra. Then, we co-add the synthetic spectra in the
same way as the science spectra (Equation~\ref{eq:co-add}),
with each pixel of the synthetic spectrum having the same weight as
the corresponding pixel in the science spectrum.

%%%%%%%%%%%%%%%%%%%%
%% Figure 3
%%%%%%%%%%%%%%%%%%%%

\begin{figure*}
 % \epsscale{0.85}
 \plotone{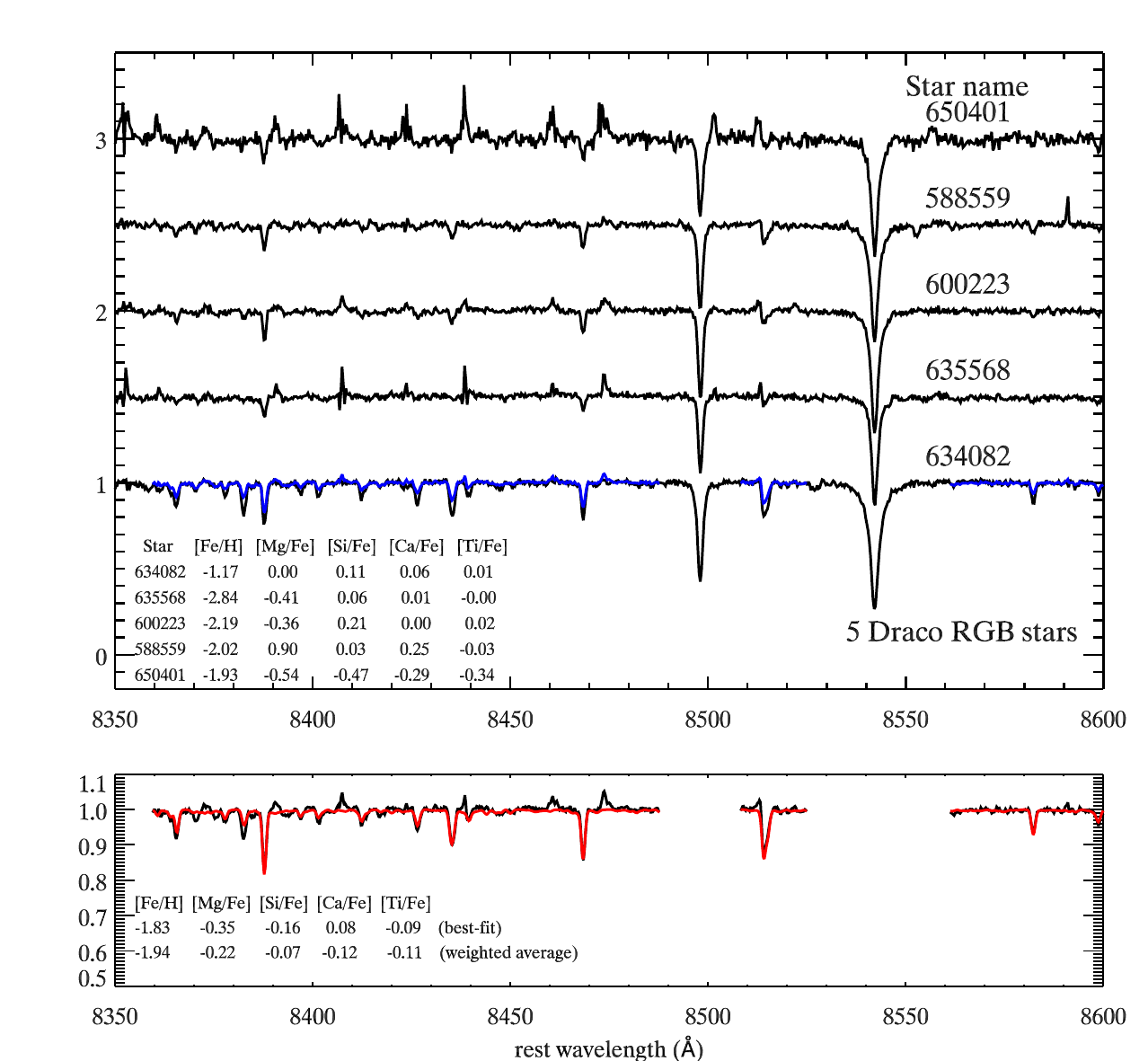}
  \caption{Top: Spectra of 5 RGB stars
     belonging to a single bin. These stars were selected to have $\log
					\it g \rm \leq 1.4 $ and are ranked by photometric [Fe/H].  Each star's individual
      spectrum is plotted in black, and the individual abundances of
      stars are shown at the bottom right.
       The co-added spectrum for these 5 stars is plotted below in blue over the 5th star for
       comparison. When we did the co-addition process, we masked
       the calcium triplet and bad regions that may bias the
       abundance measurements. Bottom: the co-added
       spectrum (black) and its best fit co-added synthetic spectrum (red).
       We fixed every star's effective temperature and surface gravity
       ($\log \it g$) by its corresponding photometric values, then
        created a grid of co-added synthetic spectra with a range in
       abundance values. The red spectrum is the best fit
       co-added synthetic spectrum for these 5 stars, the abundance of the
       best fitting synthetic spectrum and weighted average abundance
       are shown for comparison.\label{fig:spectra}}
\end{figure*}

Figure~\ref{fig:spectra} presents an example of a co-added observed spectrum (blue
in the top panel and black in the bottom panel and
its best fit co-added synthetic spectrum (red in the bottom panel). Stars with similar stellar properties
should have the most similar spectra which are then used to determine
chemical abundances and stellar parameters.  When we carried out the co-addition, we
also avoided the calcium triplet, which failed in spectra synthesis. When the comparison was executed,
only the spectral regions determined to be most sensitive to the
element were considered in the abundance determination, as showed in the
bottom panel of Figure~\ref{fig:spectra}.

\subsection{\feh\ and \afe\ Determination}

Following KGS08 and Kirby et al.\ (2009), we determine individual chemical
abundances with elemental ``masks'' that only cover the wavelength regions
that are sensitive to that particular element.  Each
element, therefore, has its own mask but, for \afe$_{\rm atm}$ the
mask is a combination of the four $\alpha$-element masks.  We adopt
the Levenberg-Marquardt algorithm implemented in IDL programs
written by Markwardt (2009) to find the best match to the co-added
observed spectra.  This algorithm finds the best co-added
synthetic spectrum by minimizing $\chi^{2}$ of pixels between the co-added
observed spectrum and the co-added synthetic spectrum over iterations. The detailed
procedure of our abundance determination is described in 11 steps:

\begin{enumerate}

\item \feh, first pass: For a bin of stars, the co-added observed
spectrum is compared with the co-added synthetic spectra. In this pass
only the spectral regions that are sensitive to Fe absorption are
considered. The \afe$_{\rm atm}$ (total $\alpha$-element abundance) and \afe$_{\rm
abund}$ (abundance of individual element, Mg, Si, Ca, Ti) were fixed at 0,
while the \feh\ value was varied until the best
fit synthetic spectrum was found. When individual synthetic spectra were
chosen, the $\it T_{\rm eff}$ and $\log \it g$ were set to be photometric values of
individual science stars in that group and the abundances were set to be
the \afe$_{\rm atm}$, \afe$_{\rm abund}$ and \feh\ values. The derived \feh\ of the best fit synthetic spectrum
was fixed and used in later steps.

\item \afe$_{\rm atm}$, first pass: All the variables were fixed
  except the total $\alpha$-element abundance, \afe$_{\rm atm}$.

\item  Continuum refinement: To refine the continuum of the observed spectrum,
the co-added science spectrum was divided by the co-added synthetic
spectrum with the parameters determined in step 1 and 2, resulting in
a quotient spectrum
without absorption lines. Then we used a B-spline with a breakpoint
spacing of 250 pixels to fit the quotient. Finally, we
divided the co-added observed spectrum by the spline fit.  This forces
a better continuum match between the observed and synthetic spectra.

\item  Step 1-3 were repeated until [Fe/H] and \afe$_{\rm atm}$
changed by less than 0.001 dex between consecutive passes.

\item  Sigma clipping: In this pass, we masked pixels whose absolute
difference from the best-fit synthetic spectrum
with the parameters determined in step 3 exceeded 2.5 times their
standard deviations. Then we used this sigma-clipped spectrum to repeat step 1.
All parameters except \feh\ were fixed. The resulting \feh\ value is
 the second pass value.

\item \afe$_{\rm atm}$, second pass: We repeated step 2 with the
sigma-clipped spectrum and the \feh\ determined from step 5.

\item \feh, third pass: Step 5 was repeated with the value of
\afe$_{\rm atm}$ determined from step 6.

\item \mgfe: Only the spectral regions sensitive to Mg absorption
lines were considered. We fixed \feh\ and \afe$_{\rm atm}$ at the values
attained from steps 7 and 6, and varied \afe$_{\rm abund}$ to measure \mgfe.

\item  \sife: We repeated step 8 but replaced \mgfe\ with \sife.

\item  \cafe: We repeated step 8 but replaced \mgfe\ with \cafe.

\item  \tife: We repeated step 8 but replaced \mgfe\ with \tife.

\end{enumerate}

\section{Comparison of co-added Spectral Abundances and Weighted Average Abundances}
The test we conduct in this study is to measure abundances on co-added
spectra and see if the results we get are consistent with the measured
abundances of the input RGB stars.  In future applications, the
individual star will be too faint to measure abundance,
so any biases should be anticipated by using this nearby star sample.
In this section, we compared the weighted-average abundances of each bin
with the abundances measured from the co-added spectra in order to
test the feasibility of the co-addition
method. The spectra we used in this test have a relatively high S/N,
and the final goal of our work is to apply this method to more distant and fainter RGB stars
beyond the Milky Way system.

\subsection{Weighted-Average Abundances}
\label{sec:waabund}

In this and subsequent sections, we refer to ``weighted-average''
abundances, which are the weighted ensemble averages of the measured
abundances for individual stars in a given bin.  Correspondingly, the
``co-added'' abundances are those derived from a measurement on the
co-added spectra.

We have tried different weights to combine the individual abundances in bins, then compared
weighted-average abundances with the abundances measured from co-added
spectra. We found that the same weights that we used in the combining of individual spectrum in co-addition (described in
Section~\ref{sec:co-add}) were the best weights to use if we wanted the
two procedures to be consistent and produce the most unbiased results.
Taking \feh\ as an example, we used the elemental mask for Fe, which covers the
wavelength regions used to determine \feh, resulting in an inverse
variance array. The average inverse variance across the entire
spectrum, calculated as in Equation~\ref{eq:maskerr} and
Equation~\ref{eq:weight1}, is then used as the
weight for that star when combining the individual abundances together
to create the weighted-average abundance for that bin.
The weights $\omega_j$ used for individual abundances in the weighted average are

\begin{equation} \label{eq:maskerr}
{\sigma^{2}_{spec,j} =
(\sum_{i=1}^{mpixel}{({\sigma_{pixel,ij}})^{-2}} \cdot M_{elemental,X} )^{-1}}
\end{equation}

\begin{equation} \label{eq:weight1}
{\omega_{j}(X) =\frac{1/{\sigma_{spec,j}^{2}}}{\sum_{j=1}^{n}{1/{\sigma_{spec,j}^{2}}}}}
\end{equation}

In Equation~\ref{eq:maskerr}, $\sigma^{2}_{pixel,ij}$ is the variance of $i$-th pixel of $j$-th spectrum in the bin.
$mpixel$ is total number of pixels in a spectrum. $M_{elemental,X}$ is the elemental mask for measurement of $X$,
  where $X$ could be abundances of \feh, \mgfe, \sife, \cafe, or \tife.
 $M_{elemental,X}$ is a binary array in which only the pixels that most sensitive to corresponding element $X$ absorption lines are set to 1.
$\sigma^{2}_{spec,j}$ is the weighted variance for the whole $j$-th spectrum.
In Equation~\ref{eq:weight1}, $\omega_j(X)$ denotes the weight of $X$ for $j$-th star, and there are $n$ stars in that group.
Then, the weighted average abundance is

\begin{equation} \label{eq:xwa}
{X_{bin,wa} =\sum_{j=1}^{n}{\omega_j(X)X_j}=\frac{\sum_{j=1}^{n}{(X_j/\sigma_{spec,j}^{2})}}{\sum_{j=1}^{n}{1/{\sigma_{spec,j}^{2}}}}}
\end{equation}

We calculated the weighted-average abundances of the four $\alpha$-elements in
the same way as \feh\ for each group. As mentioned before, some
stars' individual element abundances were unavailable and these stars were
not included in the co-added spectra, so we ignored them when
determined weighted-average abundances.

\subsection{Errors}
\label{sec:waerr}

We considered two kind of errors that contribute to the scatter in the abundance
distribution: fitting error and systematic error. The fitting error was given by the
Levenberg-Marquardt algorithm code. The MPFIT program determined the best-fit
synthetic spectrum by minimizing $\chi^{2}$ and gave an estimate of the
fitting error based on the depth of the $\chi^{2}$ minimum in parameter
space.  There are many other sources of error like the inaccuracy of atmospheric
parameters, imperfect spectral modeling and imprecise continuum
placement. We considered all other
uncertainties as systematic error. For individual stars, we used
abundance error floors derived by Kirby et al.\ (2010) as systematic
error\footnote{Table~5 of abundance error floors for five elements in
Kirby et al.\ (2010). Data used in this
work source from previous work of Kirby et al., hence, we expect the systematic errors are same for our
measurements.}. Therefore, the total error for individual stars, $\sigma_{total,j}$ for $j$-th star,
is calculated as:
\begin{equation} \label{eq:error}
{{\sigma_{total,j}}(X)=\sqrt{({\sigma_{fit,j}}(X))^2+({\sigma_{sys}}(X))^2}}
\end{equation}
where $\sigma_{fit,j}(X)$ is the fit error for abundance $X$ of the
$j$-th star, and $\sigma_{sys,j}(X)$ is systematic error.
The fit error and systematic error for individual stars should be
uncorrelated, in which case
the total error ($\sigma_{total,j}(X)$) is simply the fit error ($\sigma_{fit,j}(X)$)
and systematic error ($\sigma_{sys,j}(X)$) added in quadrature, where $X$
is either \feh\ or \afe, and $\alpha$ denotes Mg, Si, Ca, or Ti.

For the weighted average abundance $X_{bin,wa}$, we estimated the variance for each bin
weighted by $\sigma^{2}_{total,j}(X)$ which are same weights used for weighted-average abundances
(see Equation~\ref{eq:weight1} and Equation~\ref{eq:xwa}):
\begin{equation} \label{eq:errorbin}
{\sigma^{2}_{bin,wa}(X)={\omega^{2}_{j}}{\sigma^{2}_{total,j}} =\frac{{\sigma^{2}_{total,j}}/{\sigma_{spec,j}^{4}}}{({\sum_{j=1}^{n}{1/{\sigma_{spec,j}^{2}}})^{2}}}}
\end{equation}
the $\sigma_{bin,wa}(X)$, then, is the weighted error of weighted mean abundance $X_{bin,wa}$.

\begin{figure}[t]
  \epsscale{0.85}
  \plotone{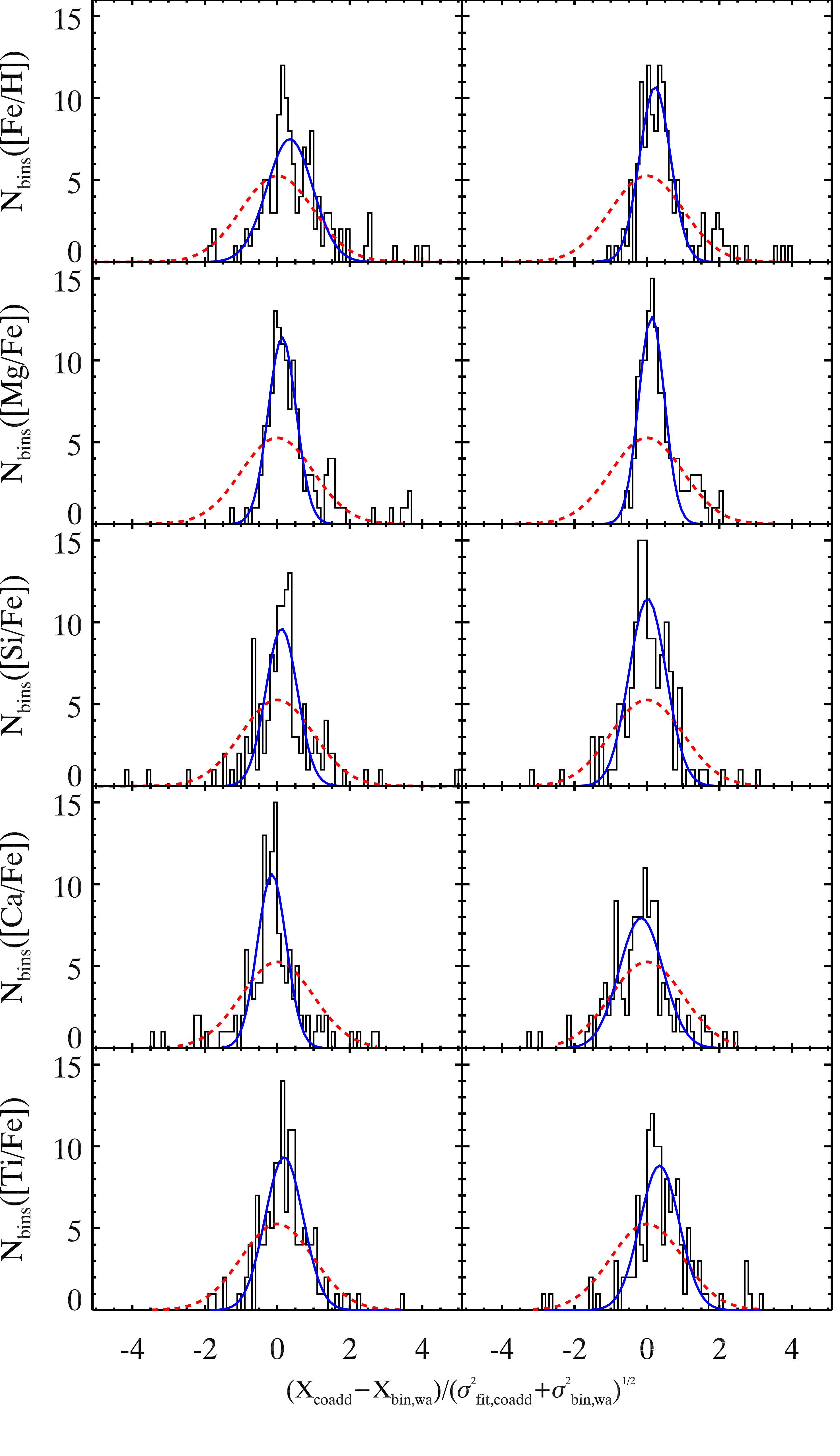}
  \caption{Distribution of the difference between the measurements from co-adds and weighted mean for 132 bins
   from 8 GCs and 8 dSphs divided by the errors of difference; systematic errors for co-adds are not included.
   Distributions in left panels are stars binned by $(\it T_{\rm
			eff})_{\rm phot}$, right panels show stars binned by $\feh_{\rm phot}$.
   The solid blue curves are the best-fit Gaussian for distributions. The dashed red curved are
   unit Gaussian with $\sigma =1$. The areas of the unit Gaussian
   are normalized to the number of bins. \label{fig:hist}}
\end{figure}

For abundances derived from the co-added spectra, we tried to use the same method in Kirby et al.\ (2010) to
estimate the systematic errors.
%Since the weighted average abundances are excellent approximations to the true values of these stars,
%we expected
The distribution of the difference between the measured co-added values and the weighted mean values
for same bins, divided by the expected errors, should be well fit by a Gaussian with unit variance, as shown in
Equation~\ref{eq:rms}:
\begin{equation} \label{eq:rms}
rms \left.\bigg(  \frac
{ {X_{coadd}}- {X_{bin,wa}}}
{ \sqrt
{({\sigma_{fit,coadd}}(X))^2+({\sigma_{bin,wa}}(X))^2+({\sigma_{sys,coadd}}(X))^2} }
 \right.\bigg)
 =1
\end{equation}
where, $\sigma_{fit,coadd}(X)$ is the fitting error of the co-added results, $\sigma_{bin,wa}(X)$ is weighted
mean error calculated from Equation~\ref{eq:errorbin}, and $\sigma_{sys,coadd}(X)$ represents the systematic error
for the co-added results. These three types of errors are supposed to
be independent to each other. In our case, however,
the fitting errors ($\sigma_{fit,coadd}(X)$) and weighted mean errors ($\sigma_{bin,wa}(X)$) are already large enough,
such that it is impossible to estimate systematic errors for the co-added
results from Equation~\ref{eq:rms}.
Figure~\ref{fig:hist} shows the distributions of the difference,
divided by the expected errors, and not including
${\sigma_{sys,coadd}}(X)$, for all GC and dSph bins. The best-fit Gaussian
is narrower than the unit Gaussian in all distributions, indicating that
either the differences of the two quantities in the numerator are too small, or
the errors in the denominator are too large. However, this does not mean we have overestimated
our uncertainties. Since we use the exact same stars for uncertainty estimating, they should contribute both to the
weighted average and to the co-added spectral measurements. It is possible that the random and systematic
 errors in the mean abundance of each bin cancel out to some
 degree. These could include errors such as those related to a spread of intrinsic abundances
within a bin, a spread caused by spectral noise, and systematic errors
resulting from $\it  T_{\rm  eff}$ mismatch. Some
of the errors are correlated between the two quantities in the numerator and cancel out
when we take the differences. On the other hand, the co-addition enhances S/N of
the spectra, then some flavors of systematic errors on the measured
abundances, e.g., those resulting the photometric estimate
of $\it T_{\rm eff}$ being different from the true $\it T_{\rm eff}$,
may indeed average down in the case of abundance determination from the
co-added spectrum. So we expect the systematic errors for co-added
abundances should be smaller than the systematic error floors of
individual stars. If the errors are correlated at some level, then
there must be a non-negligible negative covariance term in the
denominator. Both effects could be going on in our case. Here, we make a
simplification for the systematic errors of the co-added results, that we
used the abundance error floors of individual stars for the co-added
uncertainties and calculated them as:
\begin{equation} \label{eq:errco-add}
{{\sigma_{total,coadd}}(X)=\sqrt{({\sigma_{fit,coadd}}(X))^2+({\sigma_{sys}}(X))^2}}
\end{equation}
where ${\sigma_{fit,coadd}}(X)$ is the fit error of co-added abundance from Equation~\ref{eq:rms}.
$\sigma_{sys}(X)$ is the systematic error from Equation~\ref{eq:error}. ${\sigma_{total,coadd}}(X)$
 is the total error used for co-added abundances. As Figure~\ref{fig:hist} shows, our error estimates are conservative.
The true errors, accounting for covariance, must be slightly smaller. For the RGB stars of M31 satellites, there will be
non-negligible random errors that result from the analysis of low S/N spectra. Random errors often have a Gaussian normal
distribution and contribute to the total errors in the measurements. We will discuss the effects of random errors to the
error budget in future work of M31.

%%%%%%%%%%%%%%%%%%%%
%% Figure 10
%%%%%%%%%%%%%%%%%%%%

%\begin{figure*}
  \begin{figure}[t]
  \epsscale{1.2}
  \plotone{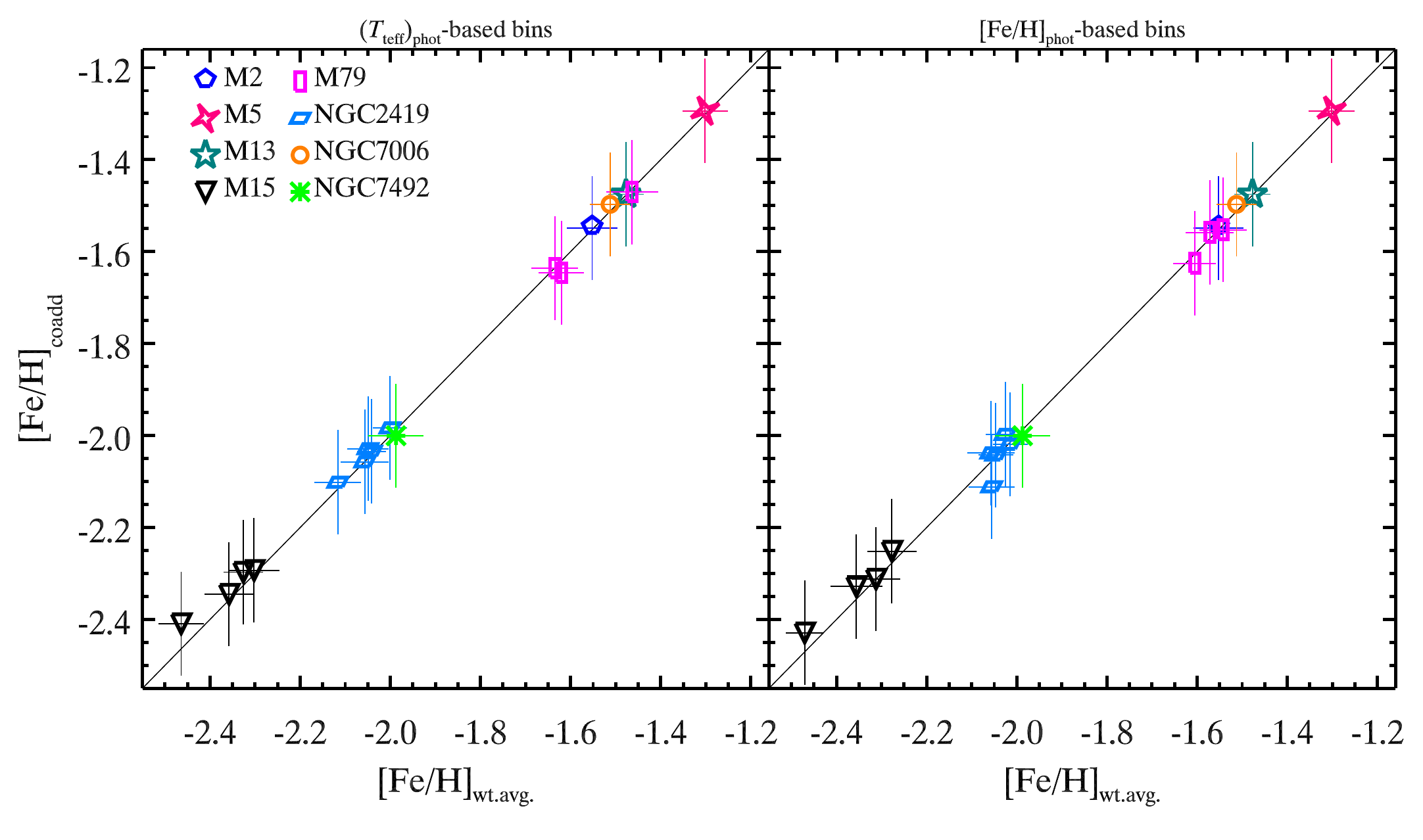}

  \plotone{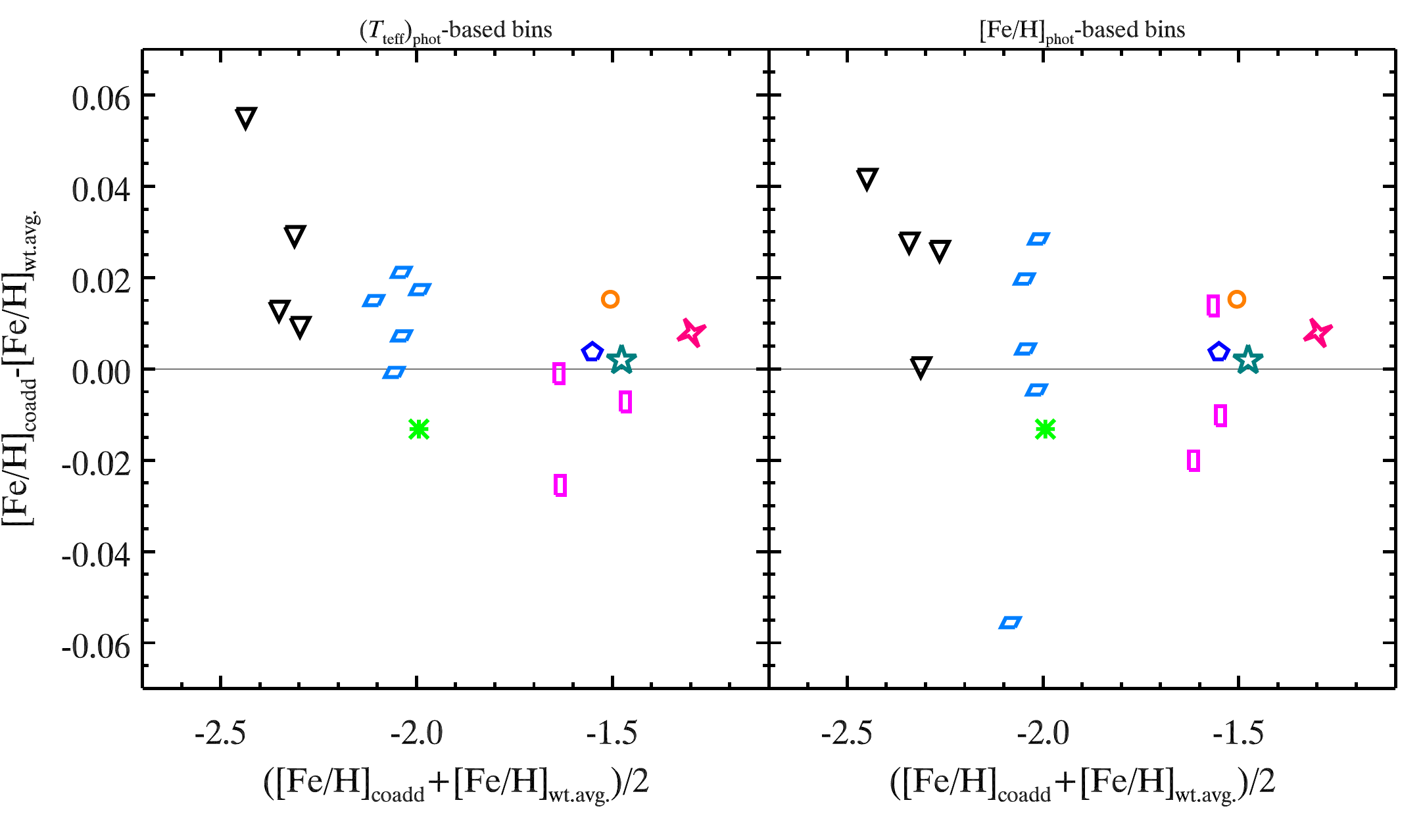}

  \caption{The comparison between weighted-average
    metallicity and metallicity from co-added spectra for 8 GCs in the Milky
    Way. Top: the value of the vertical axis represents co-added results and
    horizontal axis is weighted-average results. For weighted-average
    abundances, we measured every star's abundance and
    averaged the abundances, weighting by the elemental mask weights
				that we used in the combining of the individual spectra. We selected stars with
				$\log \it g \rm \leq 1.4$ for all 8 GCs stars and binned stars by
    photometric effective temperature (left) and photometric
    metallicity (right) respectively. Bottom: Residuals for
    co-added and weighted-average metallicity values versus their
				average. Each GC has been denoted by an unique symbol and color.
    \label{fig:gcsfeh}}
\end{figure}

%%%%%%%%%%%%%%%%%%%%
%% Figure 11 -alpha
%%%%%%%%%%%%%%%%%%%%

%\begin{figure*}
\begin{figure}[t]
  \epsscale{1.2}
  \plotone{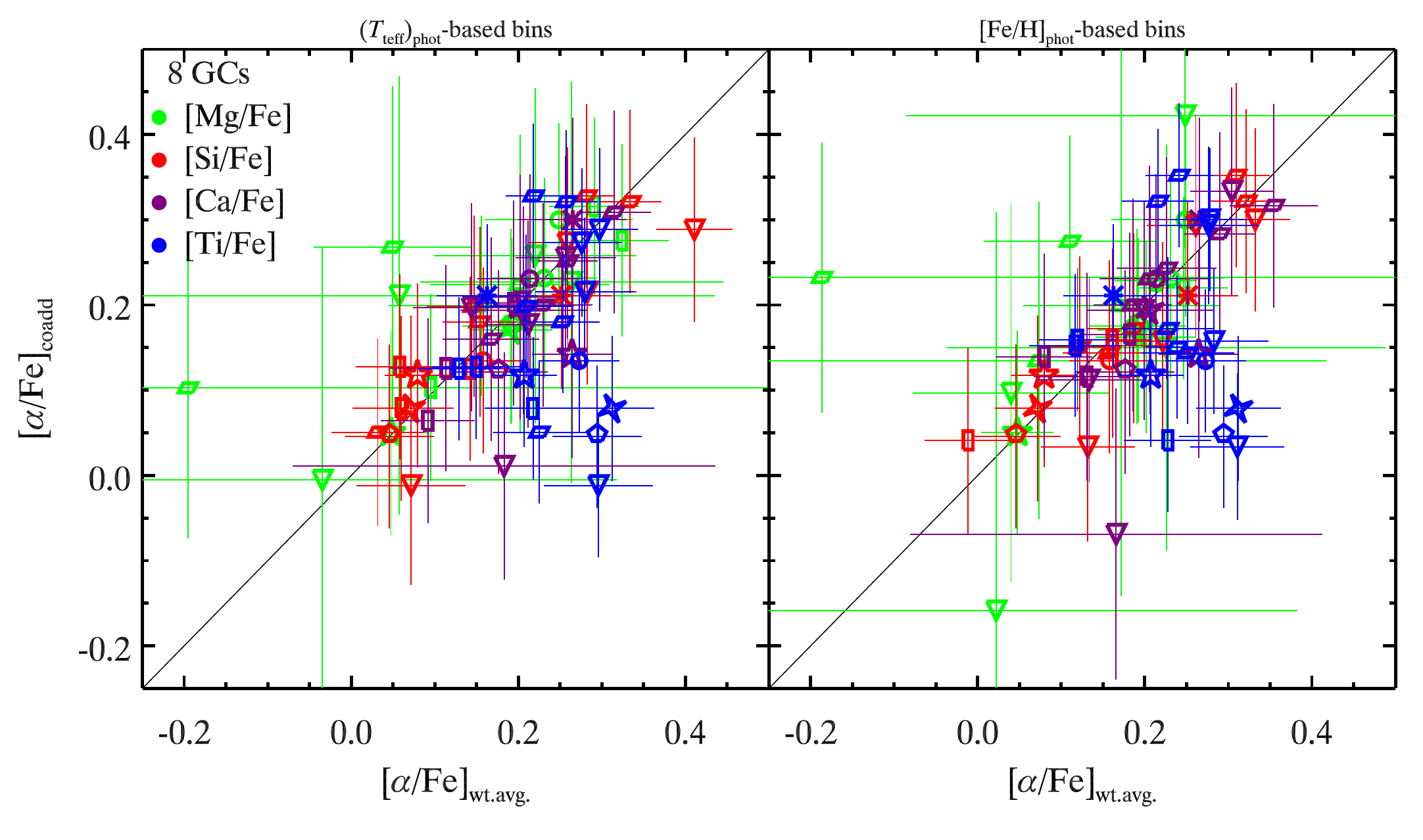}

  \plotone{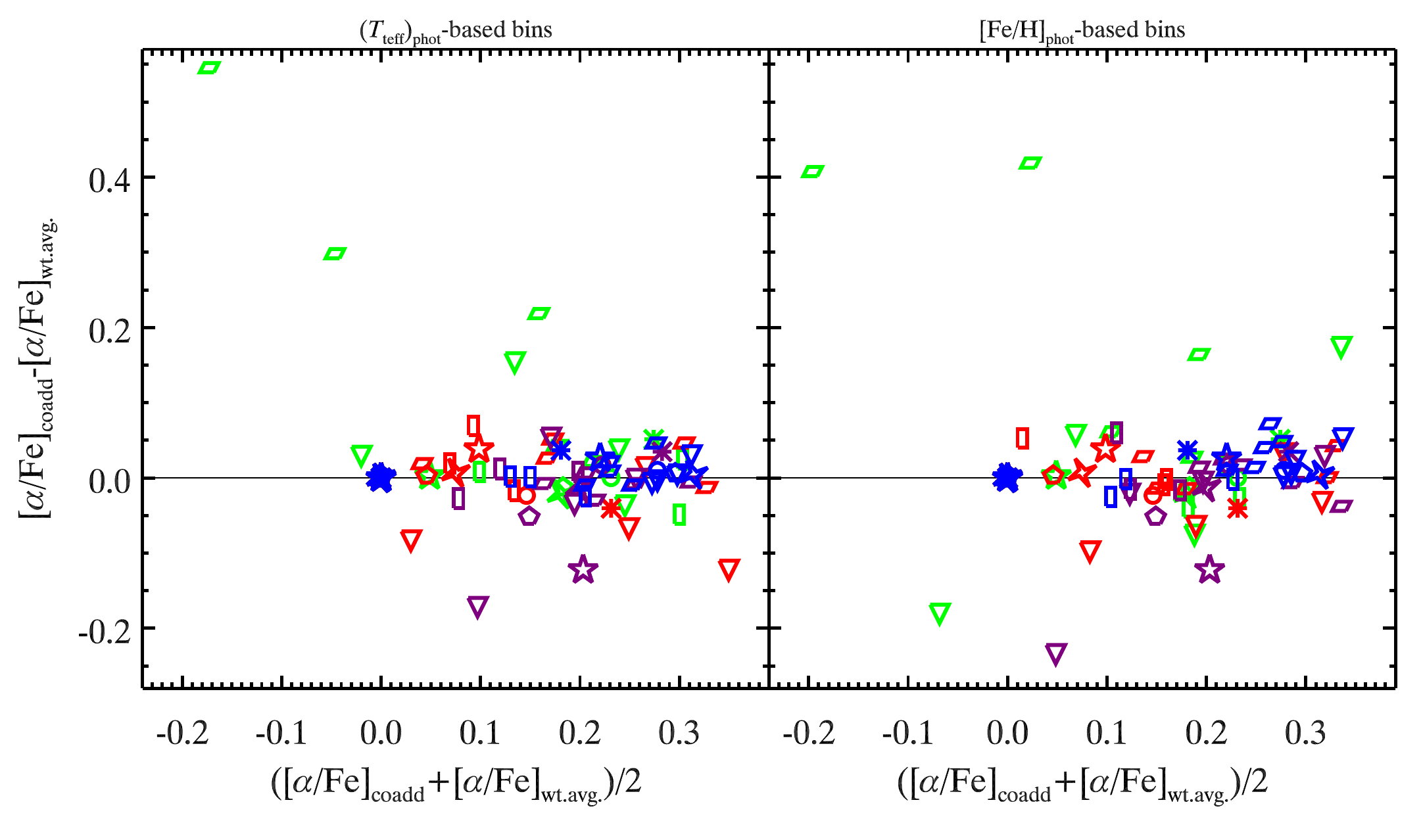}

  \caption{Same as Figure~\ref{fig:gcsfeh} except for \afe\ instead of \feh. The
		different symbols denote the different GCs as defined in Figure~\ref{fig:gcsfeh}
		but with different colors for the different
		$\alpha$-elements. \mgfe\ in green, \sife\ in red, \cafe\ in
		purple, and \tife\ in blue.
    \label{fig:gcsalpha}}
\end{figure}

\subsection{Comparison}

We expect that the abundances
measured on the co-added spectra should match
the weighted-average abundances of the stars that were used to produce
the co-add.  By using high S/N medium-resolution spectra of nearby RGB stars,
 we can know both the input and output abundances and make a
robust comparison. The co-added abundances for 8 GCs and 8 dSphs are
derived from co-added spectra as described in \S\ref{sec:co-add}, and the
weighted-average abundances are calculated based on individual
abundances from \S\ref{sec:indabund} and combined as described in
\S\ref{sec:waabund}.

For 8 GCs, the test is cleaner because we can reasonably assume no
age, metallicity, or $\alpha$ abundance
spread. Figure~\ref{fig:gcsfeh} shows the
comparisons between weighted-average metallicity and co-added spectra metallicity. Stars used here are
constrained by $\log \it g$ ($\rm \leq 1.4$) and binned by photometric
effective temperature (left figures) and photometric metallicity (right
figures). GCs are monometallic, so
the spreads in the photometric metallicity of GCs are expected to arise from
measurement uncertainty. Despite the limited number of stars in the GCs, the
two results demonstrate
a high level of agreement for these bins (top figures). The differences
between the co-added metallicities and weighted-average metallicities are less than
0.06 dex for both binned scenarios as shown in the bottom figures. In
Figure~\ref{fig:gcsalpha}, the comparisons of $\alpha$-elements on a
bin basis also match well, except \mgfe, for which the co-added
results present relatively larger values compared to the
weighted-average results, especially for
NGC2419 as shown in the two
bottom figures of residuals.

%%%%%%%%%%%%%%%%%%%%
%% Figure 4  dSph
%%%%%%%%%%%%%%%%%%%%

\begin{figure}[t]
\epsscale{1.2}
  \plotone{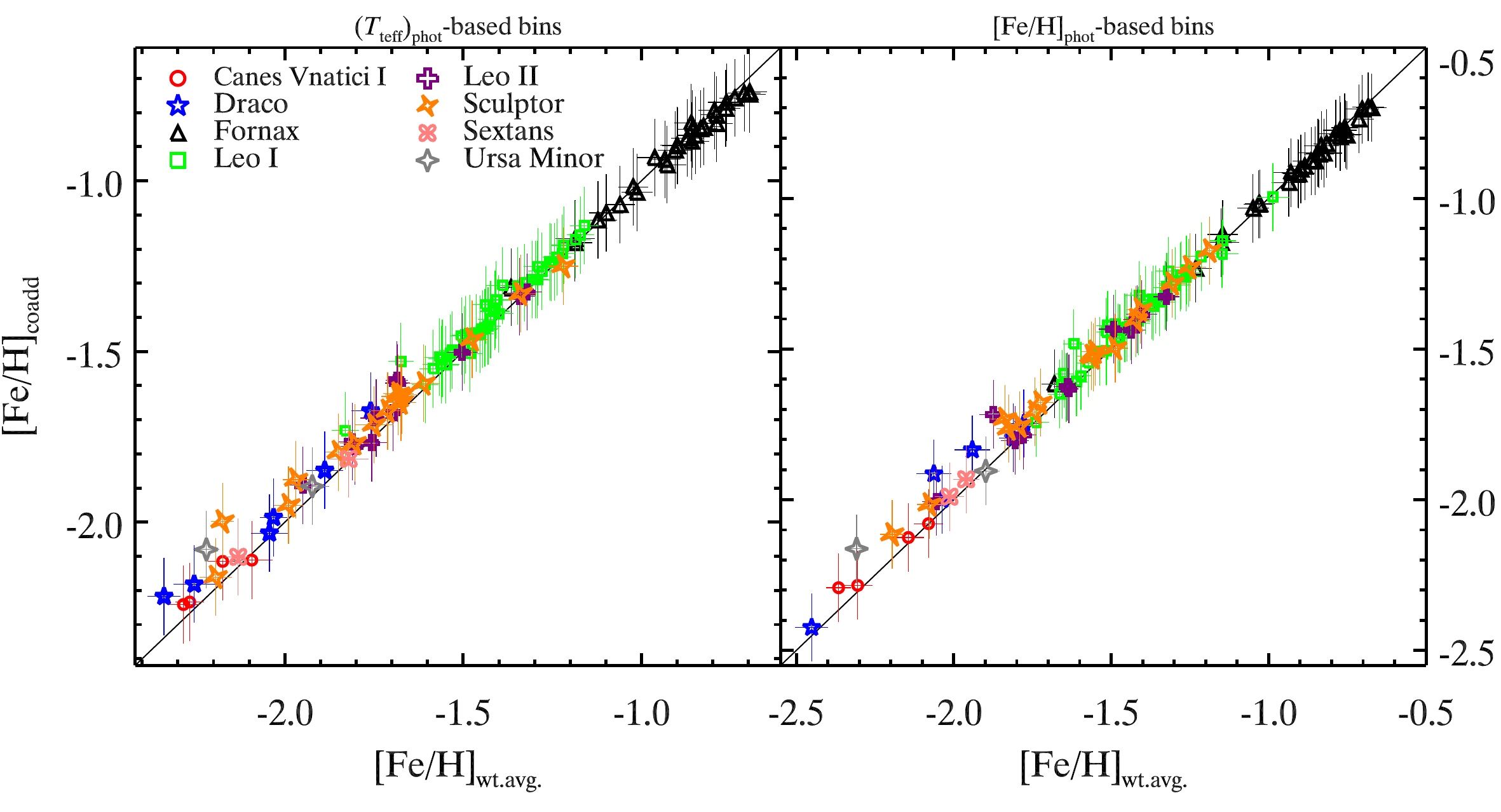}

   \plotone{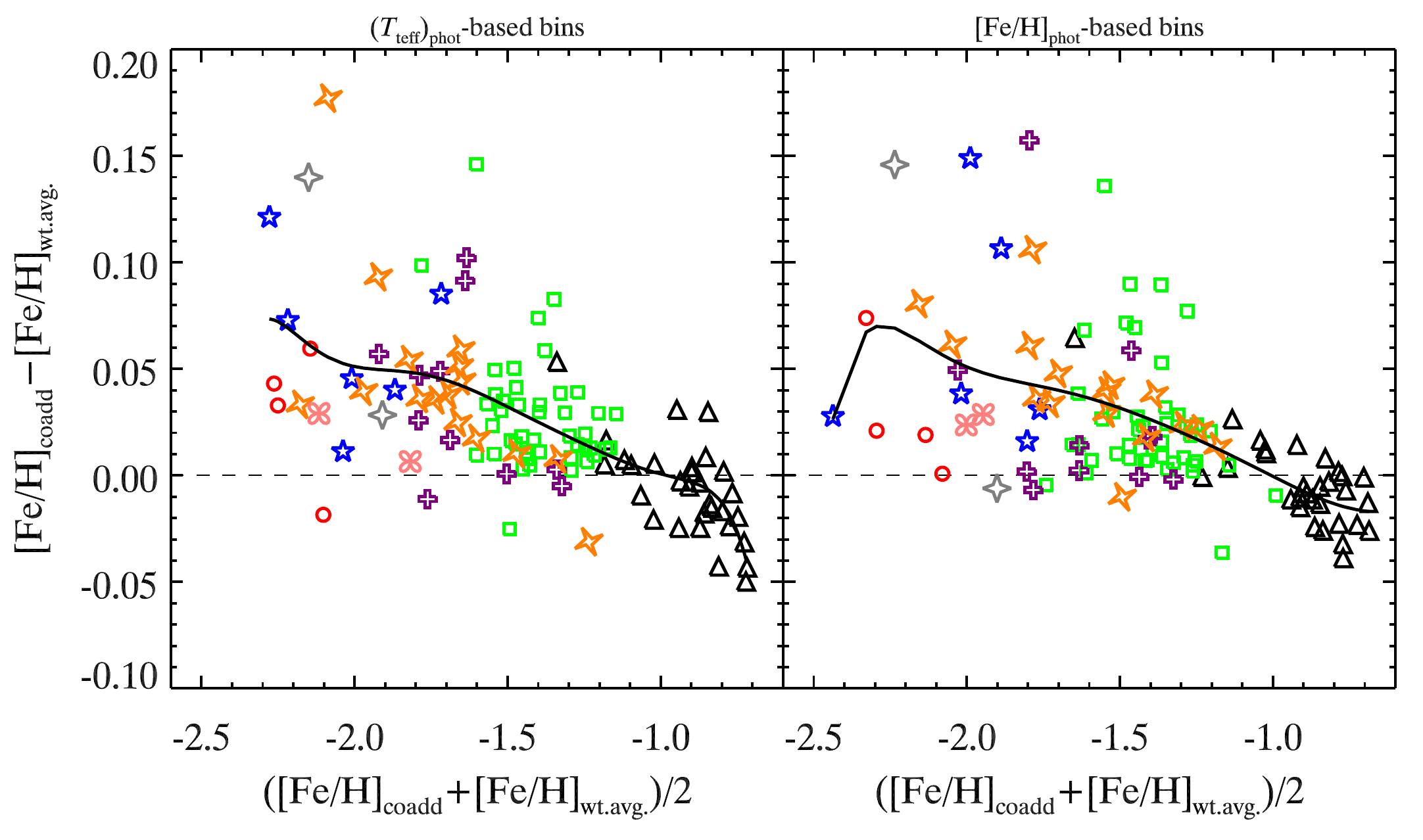}
  \caption{The comparison between weighted-average
    metallicity and metallicity from co-added spectra for 8 dSphs in the Milky
    Way. Unique symbols with different colors have been used to denote
				the 8 dSphs in the plots. Top: the value of the vertical axis represents co-added results and
    horizontal axis is weighted-average results. For weighted-average
    abundances, we measured every star's abundance and
    averaged the abundances, weighting by the elemental mask weights
				which are used in the combining of individual spectrum in co-adding
				spectrum. We selected stars with $\log \it g \rm \leq 1.4$ for all 8 dSphs stars and binned stars by
    photometric effective temperature (left) and photometric
    metallicity (right) respectively. Bottom: residuals between
    $\rm [Fe/H]_{\rm coadd}$ and $\rm [Fe/H]_{\rm wt.avg.}$ vs. the average of
    $\rm [Fe/H]_{\rm coadd}$ and $\rm [Fe/H]_{\rm wt.avg.}$. The curves
				in black are derived by polynomial
				fitting the residuals.
				\label{fig:fehcomp}}
\end{figure}

%%%%%%%%%%%%%%%%%%%%
%% Figure 5 -alpha
%%%%%%%%%%%%%%%%%%%%

\begin{figure}[t]
  \epsscale{1.2}

  \plotone{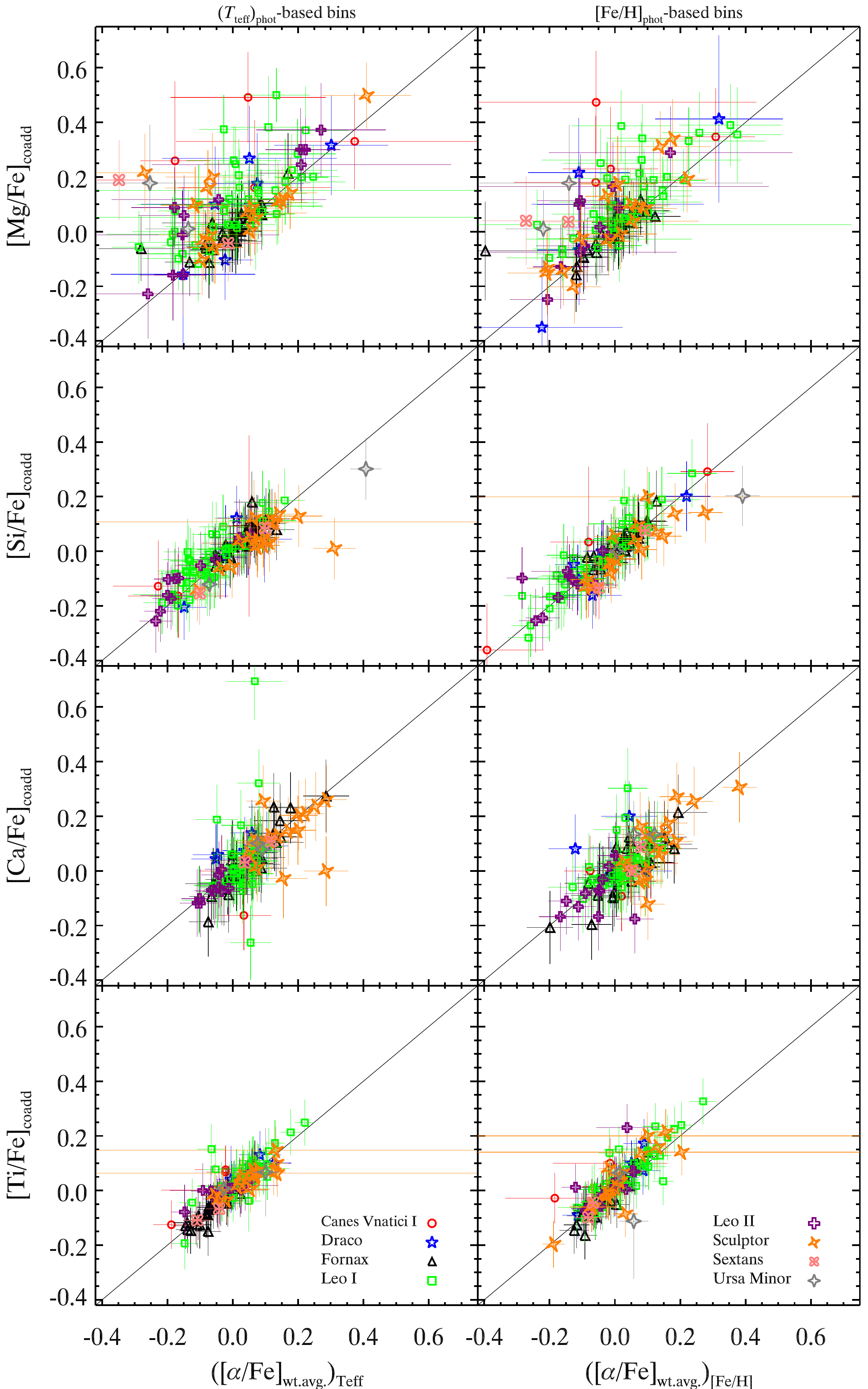}
	 \caption{Comparison for $\alpha$-element
    abundances between weighted-average abundances and co-added
    spectral abundances for four $\alpha$-elements, Mg, Si, Ca, and Ti
    (top to bottom).  Stars used for these plots are from the 8 dSph
    galaxies.  Points in the left panels are stars binned by
    $T_{\rm eff}$ and right hand panels are binned by \feh$_{\rm phot}$. The
				symbols are same as in Figure~\ref{fig:fehcomp}.
    \label{fig:alphacomp}}
\end{figure}

\begin{figure}[t]
  \epsscale{1.2}

  \plotone{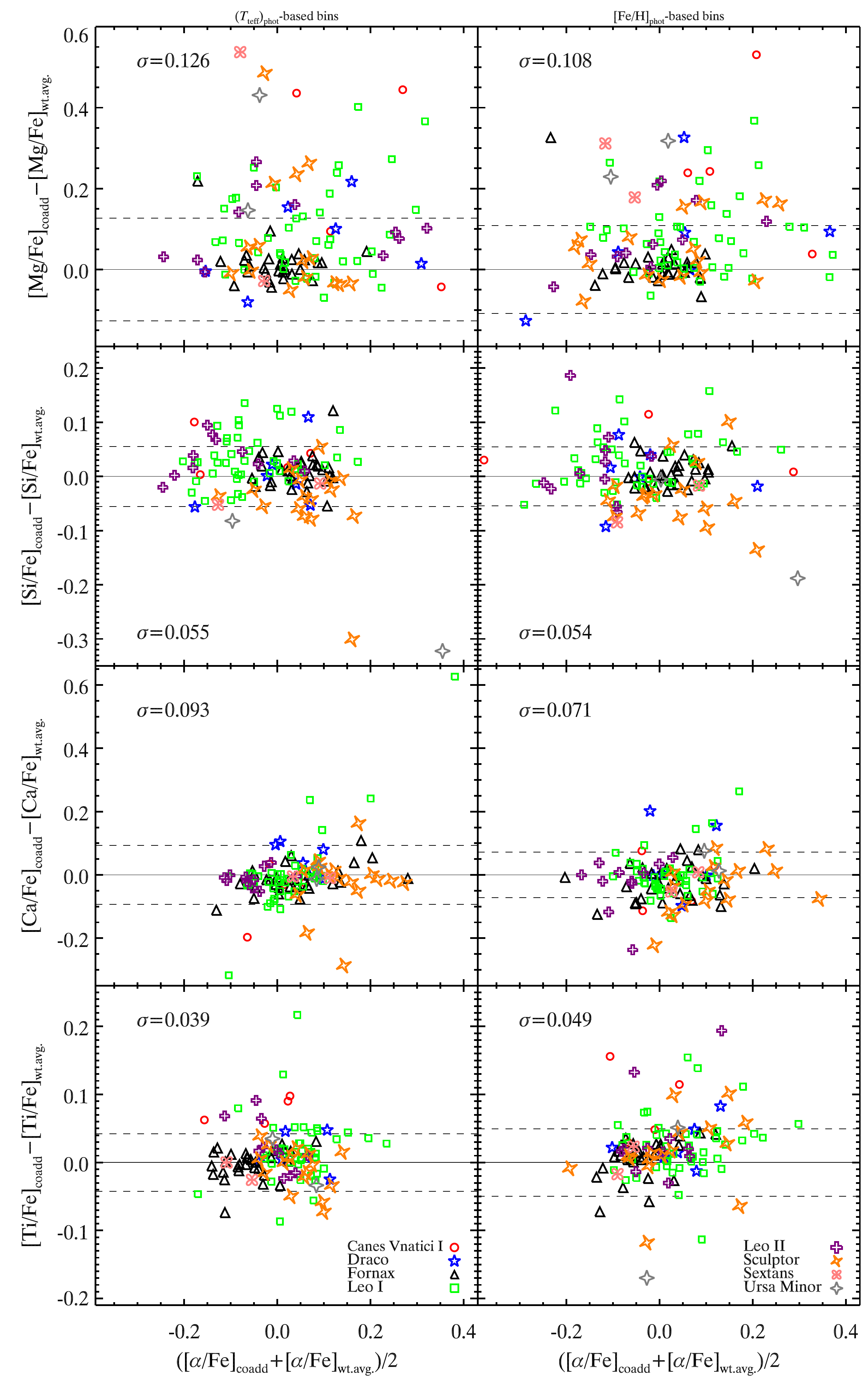}
  \caption{Comparison for $\alpha$-element
    abundances.  The difference between the weighted-average abundance and co-added
    spectral abundance for four $\alpha$-elements, Mg, Si, Ca, and Ti
    (top to bottom) versus the mean abundance.  This is the
    corresponding difference  plot for Figure~\ref{fig:alphacomp}. As
    before, points in the left panels are stars binned by
    $T_{\rm eff}$ and right hand panels are binned by \feh$_{\rm phot}$.
     We also calculate the standard deviation for
				each residual distribution draw as the dashed lines. The
				symbols are same as in Figure~\ref{fig:fehcomp}.
				\label{fig:alphadiff}}
\end{figure}

For 8 dSphs, Figure~\ref{fig:fehcomp} shows metallicity comparisons between
weighted-mean and co-added spectra measurements. The left side plots show
bins of stars grouped by photometric $\it T_{\rm eff}$ and the right side plots
show bins grouped by \feh$_{\rm phot}$. Each
point represents one bin of stars.
The two bottom plots present the difference
of the two results versus their straight averages.
The two results show a good match over $-1.7<\ \feh\ <-0.7$, for
which the difference less than 0.05~dex. For $\feh\ \lesssim-1.5$~dex,
the abundances from co-added spectra show a slight bias to be more metal-rich for both binning
approaches, and stars binned by $(\it T_{\rm eff})_{\rm phot}$ show
slightly less scatter.
We use a high-order polynomial fit to compare the bias trends of the residuals for the two binning scenarios.
There are total 115 bins for each scenario. For stars binned by photometric effective temperature, around
96 percent of bins have residuals less then 0.1~dex, and for photometric
metallicity, around 95 percent of bins have residuals less then 0.1~dex.

Figure~\ref{fig:alphacomp}  shows the comparison of four
$\alpha$-elements for 8 dSphs. Even after limiting our sample to $\log
\it g \rm \leq 1.4$, i.e., the brighter stars were chosen,
some stars still have unmeasurable elemental abundances and are not included in the bins. The plots in
Figure~\ref{fig:alphadiff} show the difference of two measurements as
in the bottom panels
in Figure~\ref{fig:fehcomp}.  The scatter here is
larger than for \feh, especially for \mgfe, for which the co-added spectra consistently
scatter to higher values.  The same trend seen in the GC stars shown in
Figure~\ref{fig:gcsalpha}.

\begin{figure}[t]
  \epsscale{1.2}

  \plotone{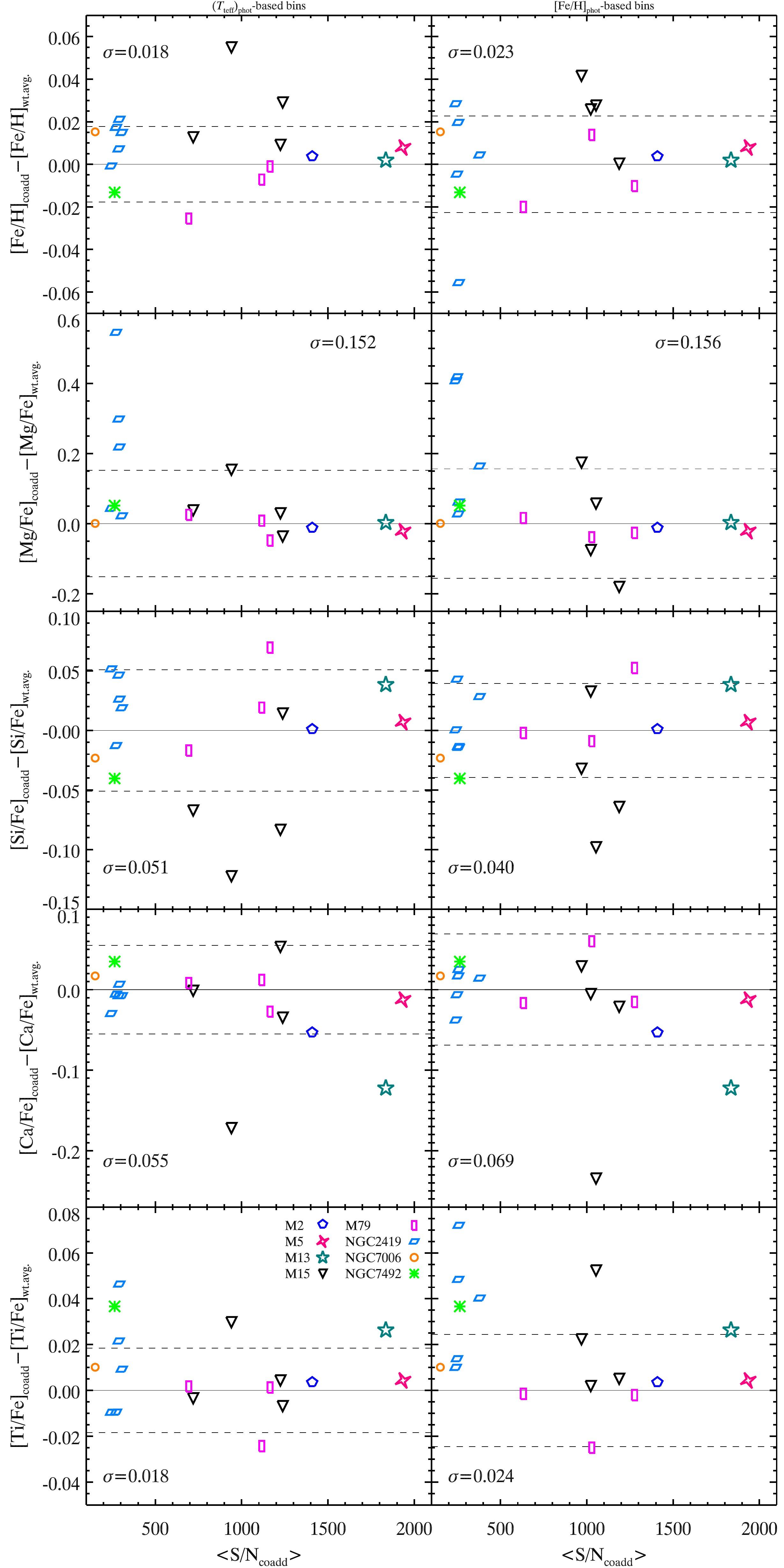}
  \caption{ The difference between abundance measurements from co-added
		spectra versus the weighted average of individual stellar abundances
		plotted versus the S/N of co-added spectra of each bin for the 8 GCs.
		The left panels use bins based on photometric estimates of stellar
		effective temperature ($\it T_{\rm eff}$)$_{\rm phot}$, while the
		right panel used bins based on photometric estimates of metallicity
		\feh$_{\rm phot}$. The dashed lines mark the range, $\pm \sigma$,
		in each panel. The symbols are the same as
		in Figure~\ref{fig:gcsfeh}. The abundance differences do not show a strong trend with
		S/N.\label{fig:sngc}}
\end{figure}

\begin{figure}[t]
  \epsscale{1.2}

  \plotone{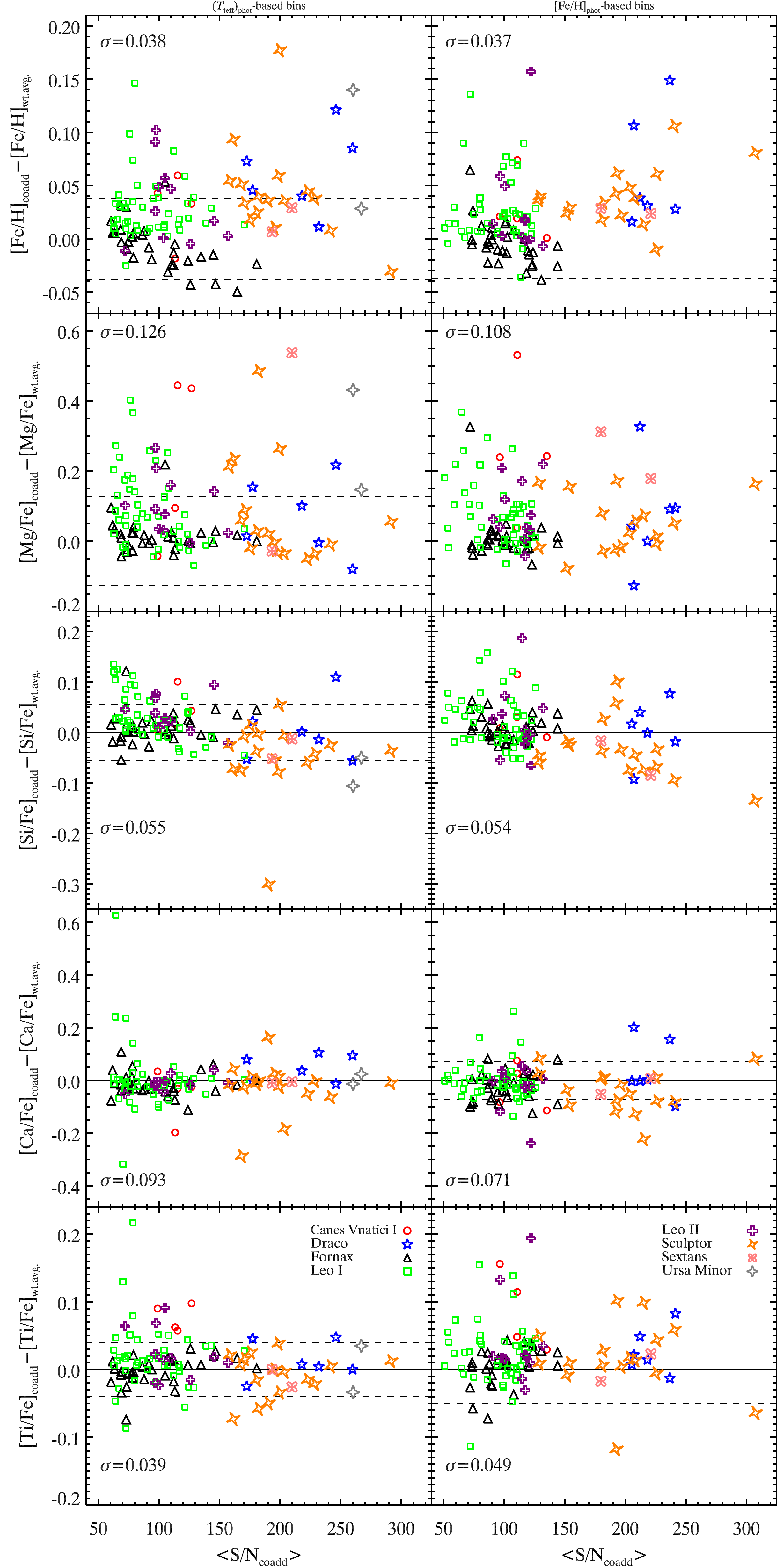}
  \caption{ The difference between abundance measurements from co-added
		spectra versus the weighted average of individual stellar abundances
		plotted versus the S/N of co-added spectra of each bin for the 8 dSphs.
		The left panels use bins based on photometric estimates of stellar
		effective temperature ($\it T_{\rm eff}$)$_{\rm phot}$, while the
		right panel used bins based on photometric estimates of metallicity
		\feh$_{\rm phot}$. The dashed lines mark the range, $\pm \sigma$,
		in each panel. The symbols are the same as
		in Figure~\ref{fig:fehcomp}. The abundance differences do not show a
		strong trend with S/N.\label{fig:sn}}
\end{figure}

There are many factors that can cause a discrepancy
between these two mean abundances, such as S/N, $\it T_{\rm eff}$ and
$\log g$, all of which may be interrelated. Since brighter RGB stars (with high S/N) often
have lower $\log \it g$ and lower $\it T_{\rm eff}$, it is
difficult to untangle which of these stellar parameters affecting the
abundance measurements most. For dSphs, there is a spread of ages and
metallicities, which removes some of the degeneracy between S/N and
$\it T_{\rm eff}$ and $\log \it g$, but our current data set still cannot sort
this out. Figure~\ref{fig:sngc} and Figure~\ref{fig:sn} show the
impact of signal-to-noise values for GCs and dSphs, respectively. In the
left panels the stars are binned by $(\it T_{\rm eff})_{\rm phot}$ and the stars are binned by $\rm
[Fe/H]_{\rm phot}$ in the right panels. The dashed lines show the standard deviations for the differences.
Here we calculate the S/N of the co-added spectrum weighted by the
inverse variance:			
\begin{equation} \label{eq:co-addsn}
{\langle\rm S/N_{\rm coadd}\rangle=\frac{\sum_{i=1}^{n}{(1/\sigma_{pixel,i}) \cdot
(1/\sigma_{pixel,i}^{2})}}{\sum_{i=1}^{n}{1/{\sigma_{pixel,i}^{2}}}}}
\end{equation}
where $\rm S/N_{ \rm coadd}$ is the signal-to-noise ratio for co-added spectrum,
$\sigma_{pixel,i}$ is the error of the $i$-th pixel of co-added spectrum.
GCs have much higher S/N for their co-added spectra compared to dSphs. For dSph stars binned by
$(\it T_{\rm eff})_{\rm phot}$, the discrepancies are slightly pronounced with S/N decreasing,
 and seem to be more concentrated compared with the stars binned
by $\rm [Fe/H]_{\rm phot}$ in
Figure~\ref{fig:sn}. In general, however, the abundances are in good
agreement for both binning methods, within 0.1~dex for GCs
and within 0.2~dex for dSphs, with \mgfe\ having the most outliers.

%%%%%%%%%%%%%%%%%%%%
%% Figure 6
%%%%%%%%%%%%%%%%%%%%

%%%%%%%%%%%%%%%%%%%%
%% Figure 6
%%%%%%%%%%%%%%%%%%%%

%%%%%%%%%%%%%%%%%%%%
%% Figure 7
%%%%%%%%%%%%%%%%%%%%

\begin{figure}[t]
  \epsscale{1.2}

  \plotone{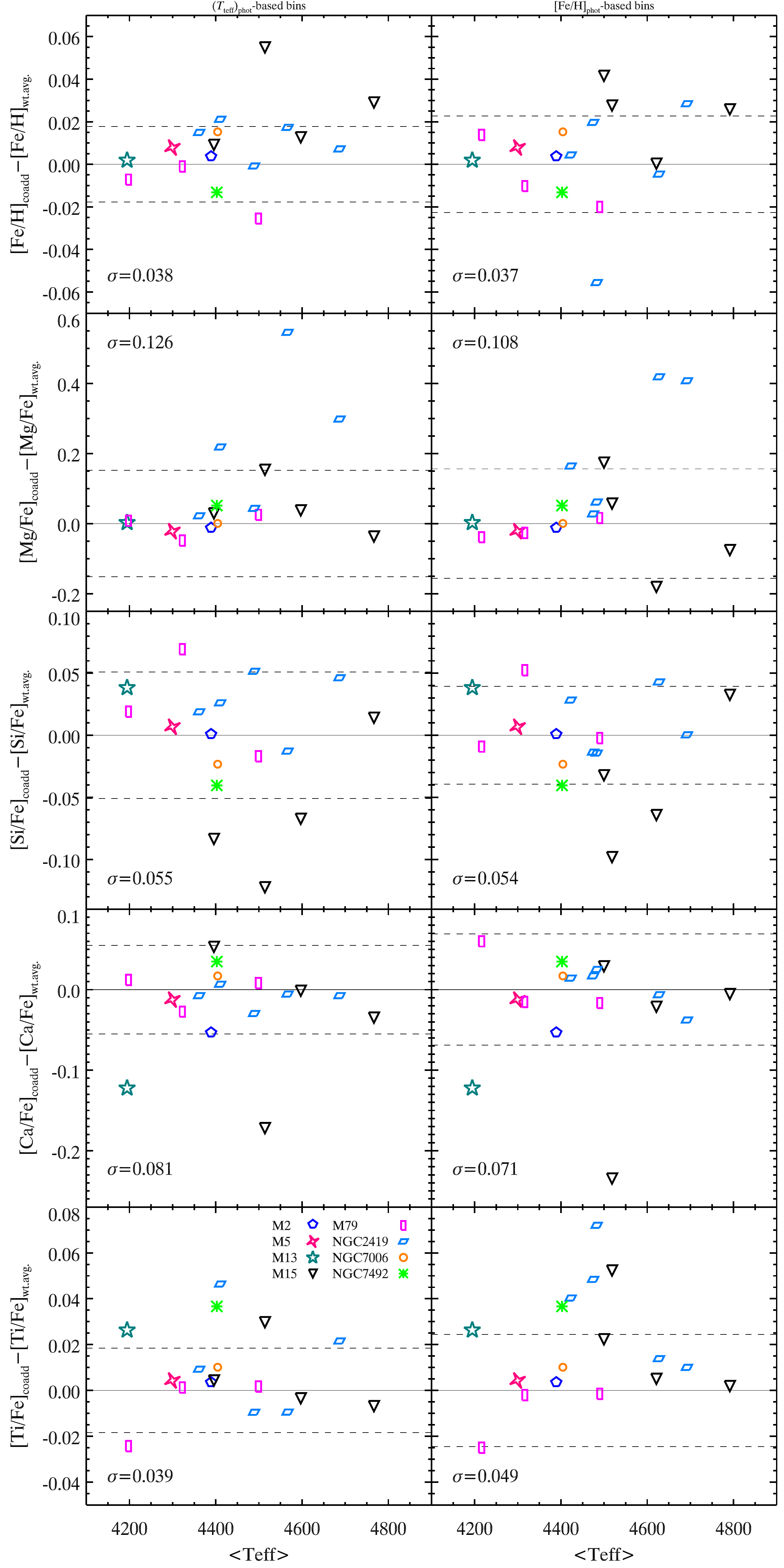}
  \caption{Same as Figure~\ref{fig:sngc} except the abundance differences
		are plotted versus mean of stellar effective temperature for each bin. The
				symbols are the same as in Figure~\ref{fig:gcsfeh}.\label{fig:teffgc}}
\end{figure}

%%%%%%%%%%%%%%%%%%%%
%% Figure 7
%%%%%%%%%%%%%%%%%%%%

\begin{figure}[t]
  \epsscale{1.2}

  \plotone{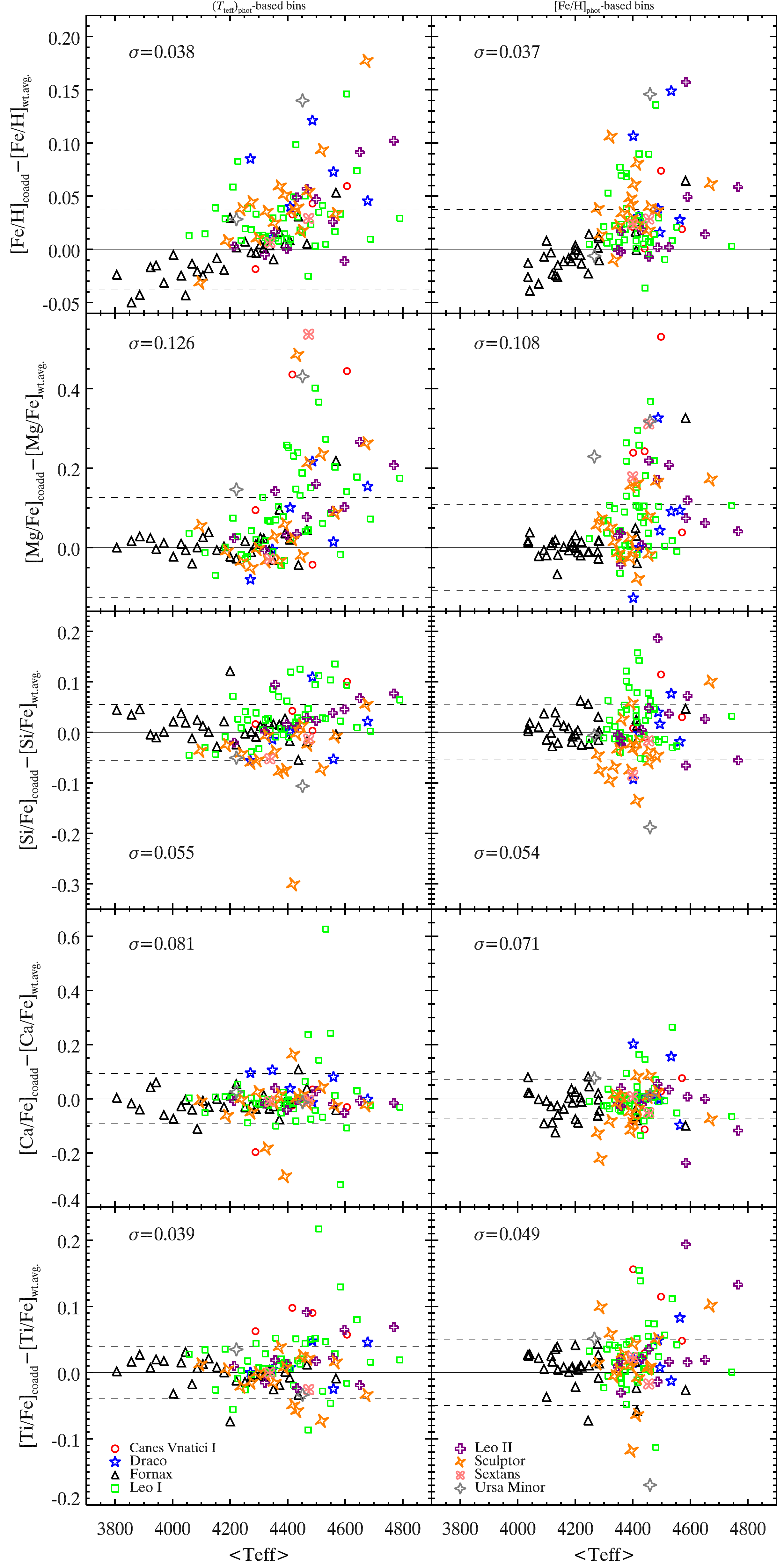}
  \caption{Same as Figure~\ref{fig:sn} except the abundance differences
		are plotted versus mean of stellar effective temperature of each bin. The differences tend
		to be larger and biased toward positive values at higher
		$\it T_{ \rm eff}$ values especially for \feh\ and \mgfe. The
				symbols are the same as in Figure~\ref{fig:fehcomp}.\label{fig:teff}}
\end{figure}

Figure~\ref{fig:teffgc} and Figure~\ref{fig:teff} consider the impact of
$\langle \it T_{\rm eff} \rangle$ for
the abundance discrepancies. For each bin, We combined the individual $\it T_{\rm
eff}$ weighted by the total error of the star to calculate the
mean $\langle \it T_{\rm eff}\rangle$.
 The left panels show stars grouped by $(\it T_{\rm eff})_{\rm phot}$ while the right
panels show stars binned by \feh$_{\rm phot}$.
The dashed lines represent one standard deviation. There is no obvious trend
shown in Figure~\ref{fig:teffgc} for GCs, but for
dSphs, there appears a slight bias in \feh, where \feh$_{\rm coadd}$ is lower at lower
$\langle \it T_{\rm eff}\rangle $.  Also, the differences in \mgfe\ are markedly higher when
$\rm 4400K \leq \langle \it T_{\rm eff}\rangle \leq \rm 4600 K$ for both binning approaches.

\begin{figure}
  \epsscale{1.2}

  \plotone{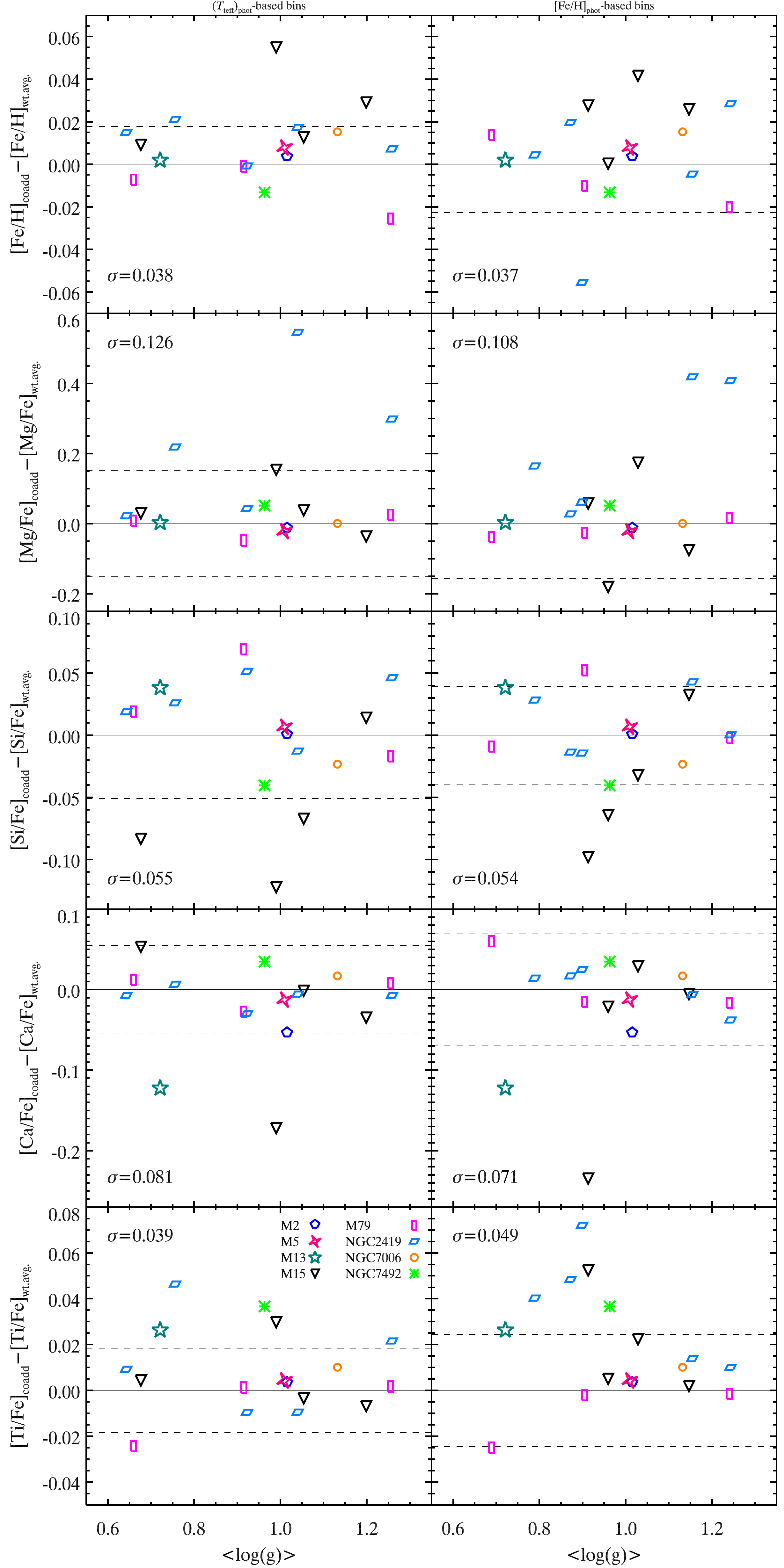}
  \caption{Same as Figure~\ref{fig:sngc} except the abundance differences
		are plotted versus mean of stellar surface gravity for each bin. The
				symbols are same as in Figure~\ref{fig:gcsfeh}.\label{fig:loggc}}
\end{figure}

%%%%%%%%%%%%%%%%%%%%
%% Figure 8
%%%%%%%%%%%%%%%%%%%%

\begin{figure}
  \epsscale{1.2}

  \plotone{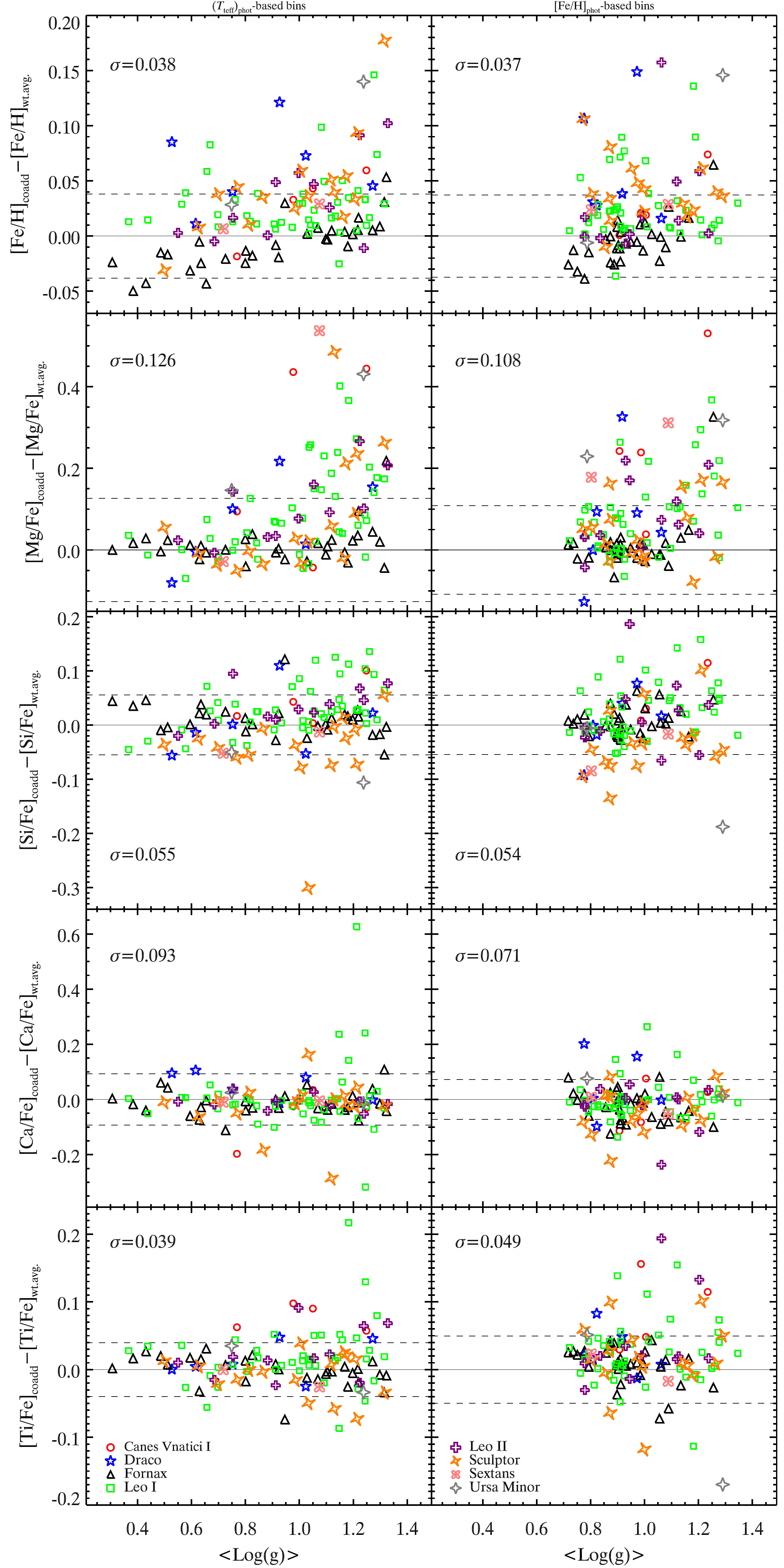}
  \caption{Same as Figure~\ref{fig:sn} except the abundance differences
		are plotted versus mean of stellar surface gravity for each bin. The differences tend
		to be larger and biased toward positive values at higher
		$\log \it g$ values especially for \feh\ and \mgfe. The
				symbols are same as in Figure~\ref{fig:fehcomp}.\label{fig:logg}}
\end{figure}

Figure~\ref{fig:loggc} and Figure~\ref{fig:logg} consider the impact of
$\langle \log \it g  \rangle$ for the abundance discrepancies.
 The photometric $\langle \log \it g \rangle$ for each bin is derived by combining individual values weighted with total
error of star. Like $\langle \it T_{\rm eff} \rangle$, the differences in \mgfe\
are tied to $\langle \log \it g \rangle $, particularly for stars binned by
$\it T_{\rm eff}$ with larger scatter for $\langle \log \it g \rangle \rm >0.9$.
The discrepancies in metallicities and $\alpha$-elements
for GCs seem to be not very sensitive to $\rm \langle S/N _{\rm
coadd}\rangle $, $\langle \it T_{\rm eff}\rangle$, or $\langle \log
\it g \rangle$, and we suspect
the limited number of bins (only 17 bins for GCs) is one reason.

\subsubsection{Photometric metallicity}
\begin{figure}[t]
\epsscale{1.2}
  \plotone{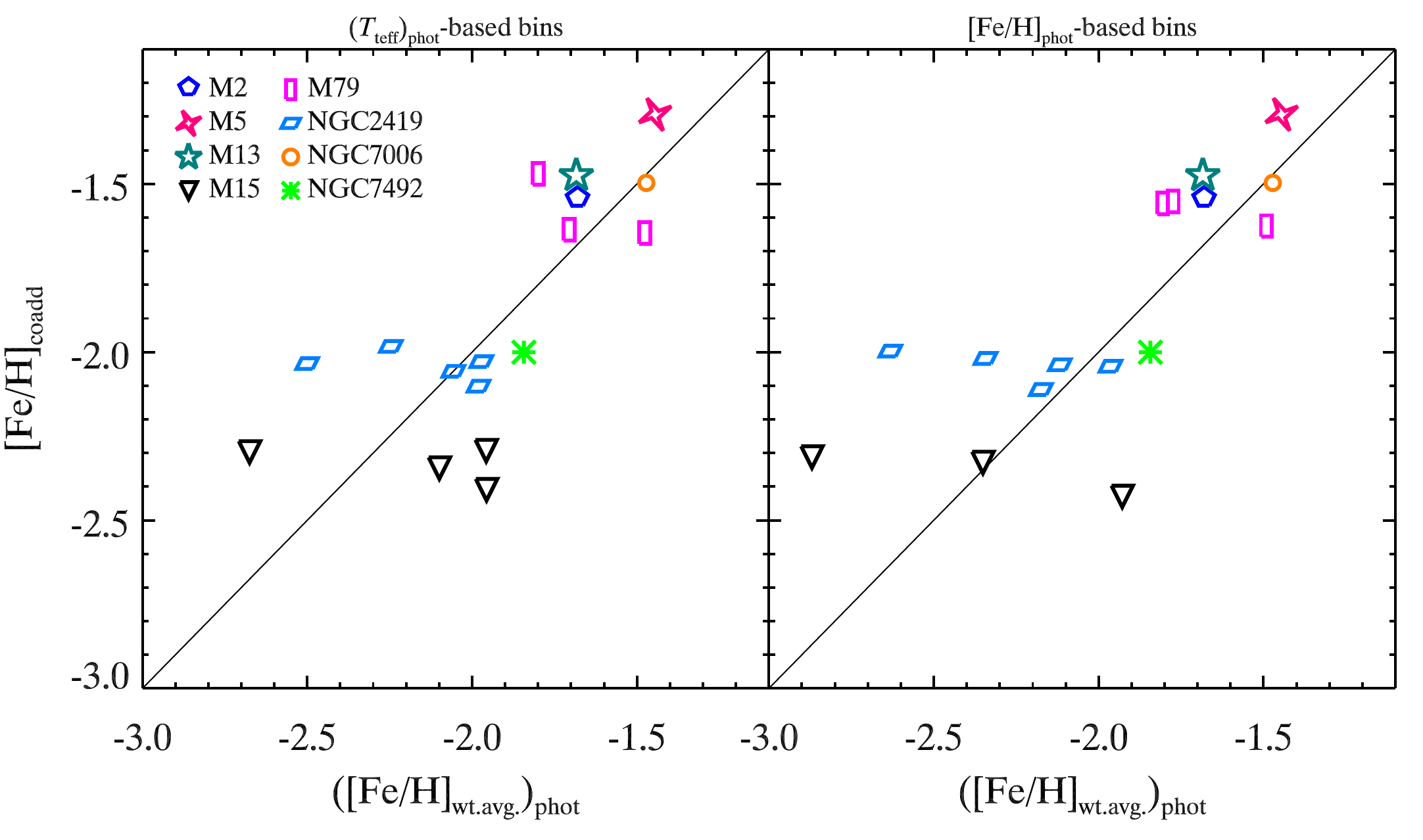}
  \plotone{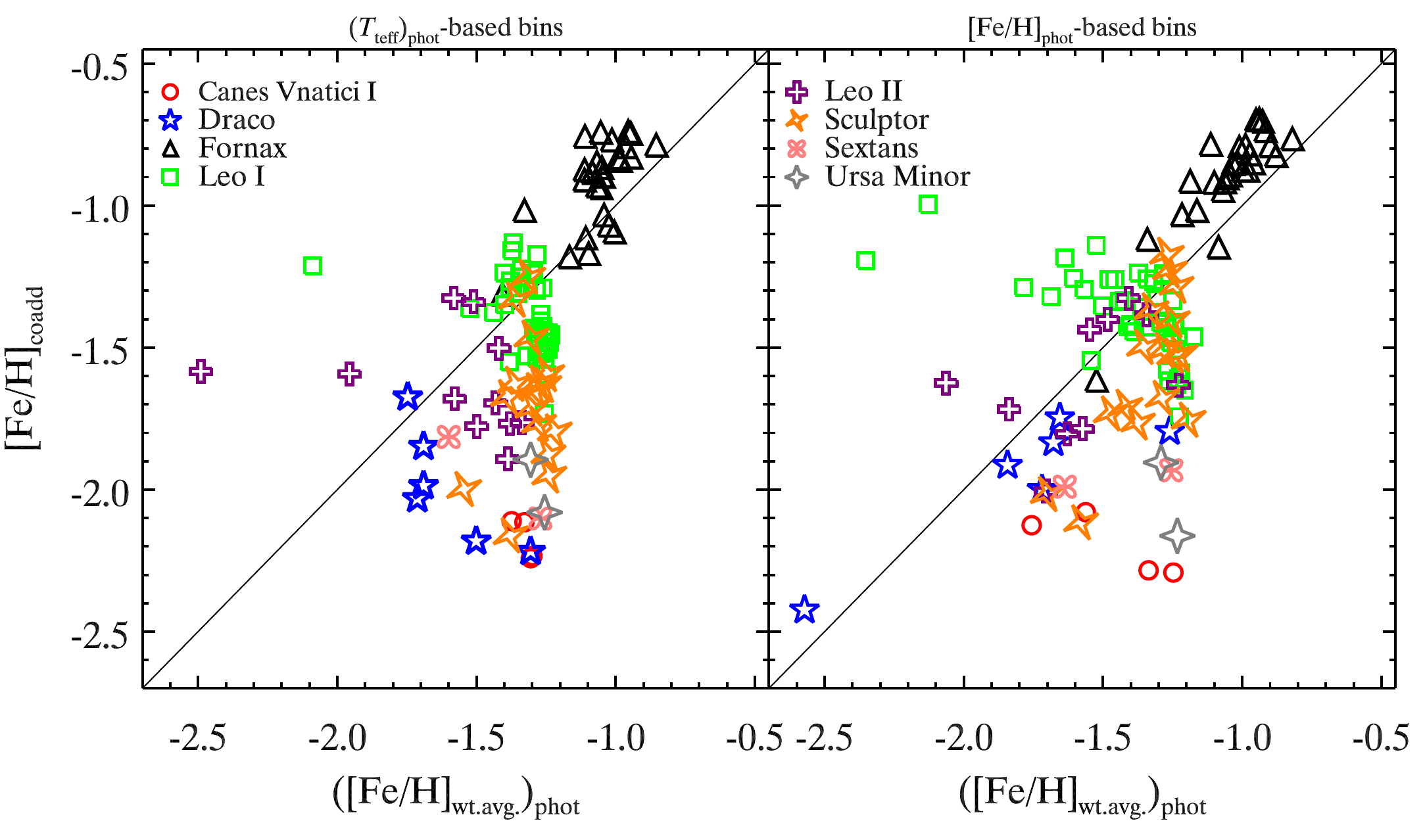}
  \caption{The comparison between weighted-average
    photometric metallicity and metallicity from co-added spectra for
				8 GCs (top) and 8 dSphs (bottom) in the Milky
    Way. The value of the vertical axis represents co-added results and
    horizontal axis is weighted-average results. For weighted-average
    abundances, we measured every star's metallicity by
				isochrone-fitting with Yonsei-Yale theoretical isochrones. We
				assumed an age of 14 Gyr for all RGB stars and \afe$ =+0.3$. Then
				we averaged photometric metallicity for each bin weighting by
				photometric errors. Here all stars satisfy $\log \it g \rm \leq
				1.4$. Stars are binned by photometric effective temperature (left) and photometric
    metallicity (right) respectively. The symbols are same as in
				Figure~\ref{fig:gcsfeh} for GCs and Figure~\ref{fig:fehcomp} for
				dSphs. One bin of M15 has been ignored in the top-right panel whose weighted-average photometric
                                metallicity is less than $-3.0$.
				\label{fig:fehcompphot}}
\end{figure}

In order to make a robust comparison, we also compared metallicity
derived from co-added spectra ($\feh_{ \rm coadd}$) to the weighted
average of photometric metallicity ($\feh_{ \rm phot}$) as shown in
Figure~\ref{fig:fehcompphot}. We combined the photometric metallicities
of member stars in each bin weighted by their photometric errors. In
the top panels of Figure~\ref{fig:fehcompphot}, the photometric
metallicities of weighted average bins present scatter for M15 and
NGC2419, but this is still acceptable given the uncertainty in photometric
measurements. NGC7006 shows the best agreement for both binning
methods. For dSphs, as shown in the bottom panels of
Figure~\ref{fig:fehcompphot}, we roughly get similar results as in
Lianou et al. (2011), in which mean the photometric metallicity estimate only has
limited reliability. There is a significant scatter in the dSphs, except for Fornax which
seems to show a better agreement and one possible reason is the choice of photometric filters.
In our data set, the photometry for Fornax is in the B and R bands.  Almost all of the
other galaxies have photometry in V/I or $\rm M/T_2$. The B band is more
metallicity sensitive than the V or M bands. Lianou et al. (2011) only
considered V/I photometry. Therefore, it's possible a large amount of
the uncertainty in $\feh_{\rm phot}$ comes from a poor choice of
filters. We will discuss the impact of our age assumption in \S4.4.4.

\subsubsection{Degradation of spectra}
\begin{figure}[t]
  \epsscale{1.2}

  \plotone{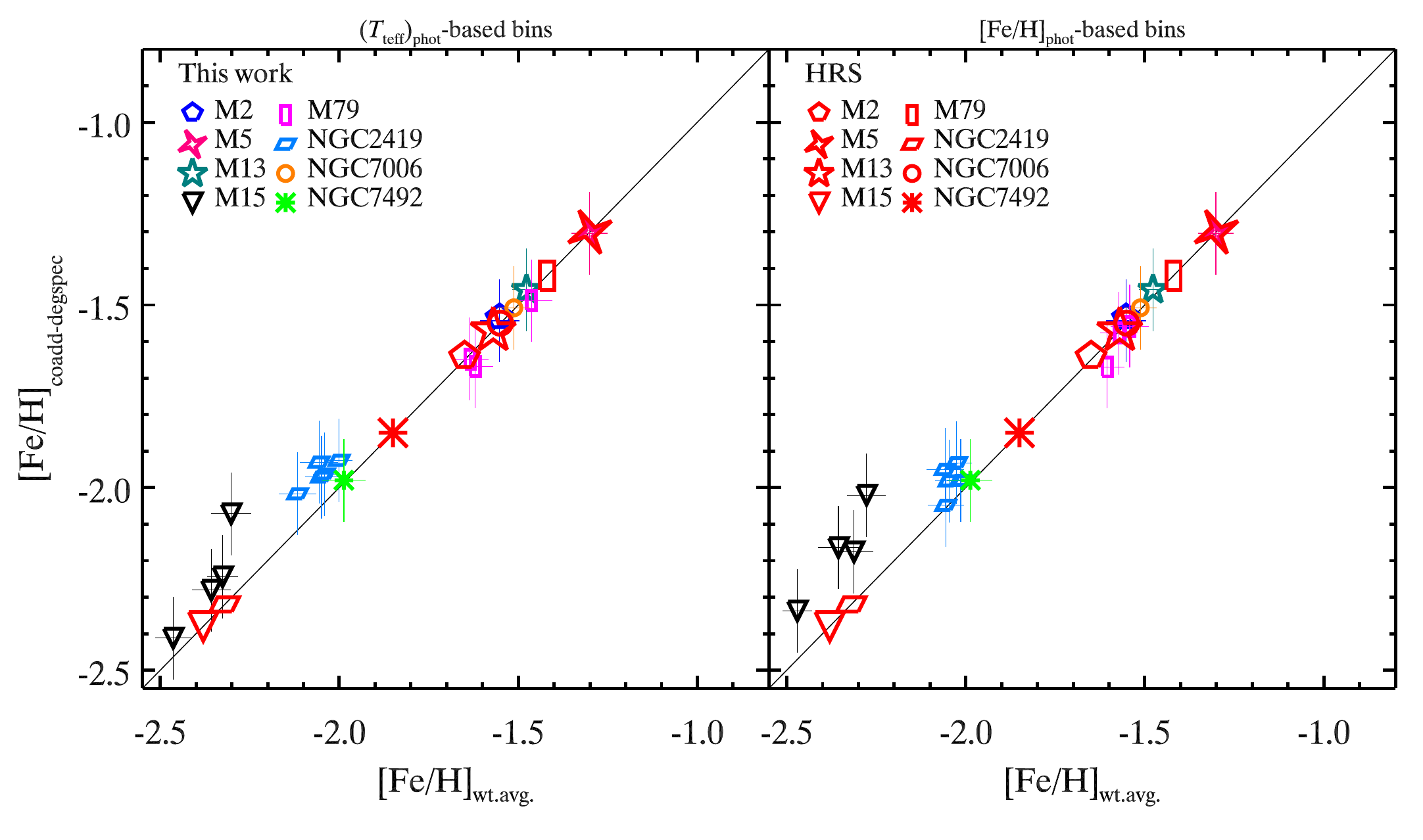}
	 \caption{The comparison between weighted-average metallicity and metallicity from co-added degraded
    spectra for 8 GCs in the Milky Way. The value of the vertical axis represents co-added results and horizontal
    axis is weighted-average results. For weighted-average abundances, we measured every star's abundance and
    averaged the abundances, weighting by the elemental mask weights which we used in the combining of individual
    spectrum. We selected stars with $\log \it g \rm \leq 1.4$
    for all 8 GCs stars and binned stars by photometric effective temperature (left) and photometric
    metallicity (right) respectively. The symbols are same as in Figure~\ref{fig:gcsfeh}. The symbols in red are the metallicities measured
    by high-resolution spectroscopy (HRS). M2 (NGC7089): Harris catalog (http://physwww.physics.mcmaster.ca/~harris/mwgc.dat).
    M5 (NGC5904), M13 (NGC6205), M15 (NGC7078), M79 (NGC1904), NGC2419, NGC7006, and NGC7492: Pritzl et al. (2005).
    Figure~\ref{fig:gcsfeh} presents the same information based on non-degraded spectra with the observed S/N.
    \label{fig:deggc}}
\end{figure}

\begin{figure}[t]
  \epsscale{1.2}

  \plotone{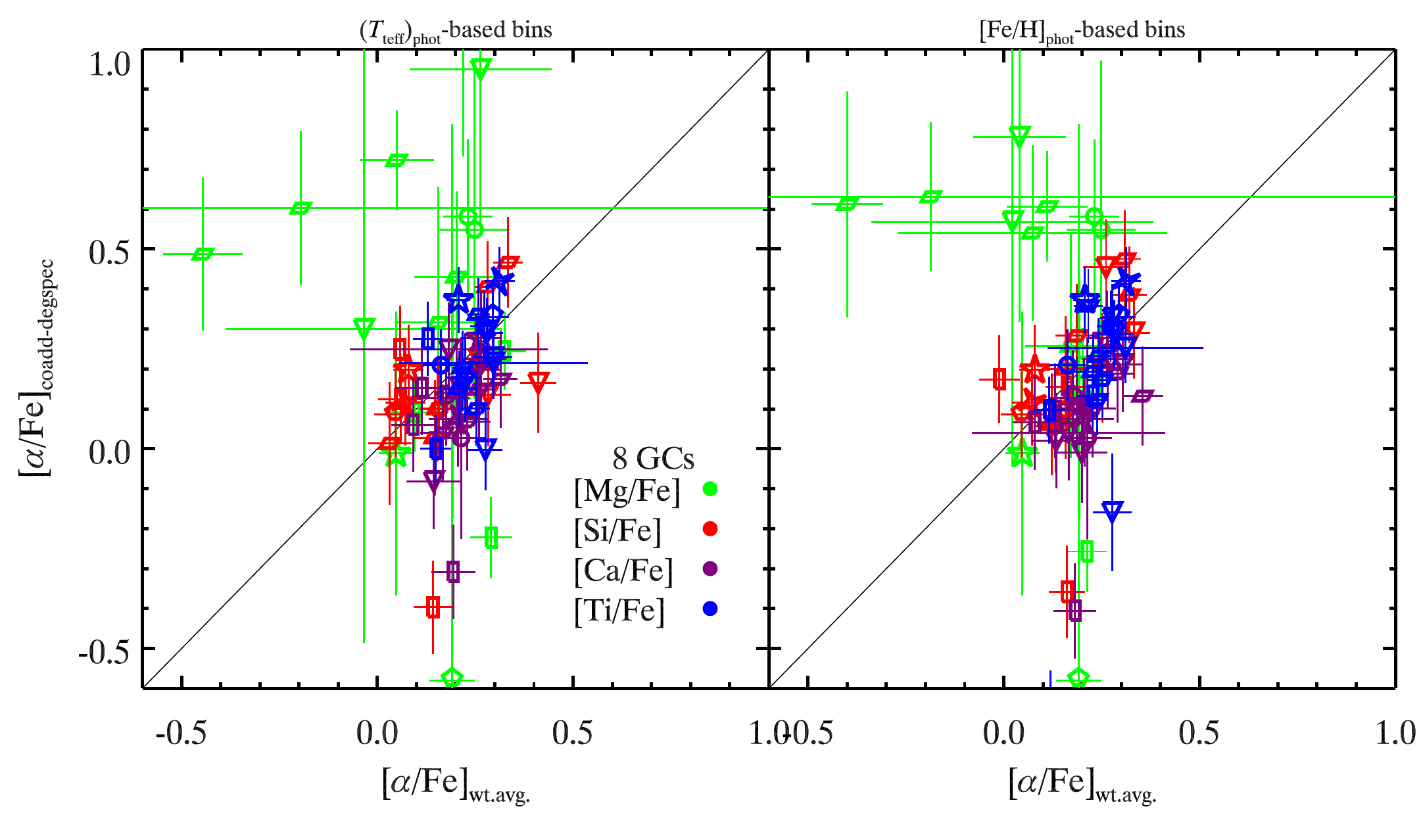}
	 \caption{Same as Figure~\ref{fig:deggc} except for \afe\ instead of \feh. The different symbols denote
    the different GCs as defined in Figure~\ref{fig:deggc} but with different colors for the different $\alpha$-elements.
    \mgfe\ in green, \sife\ in red, \cafe\ in purple, and \tife\ in blue. Figure~\ref{fig:gcsalpha} presents
    the same information based on non-degraded spectra with the observed S/N.
    \label{fig:deggca}}
\end{figure}

\begin{figure}[t]
  \epsscale{1.2}

  \plotone{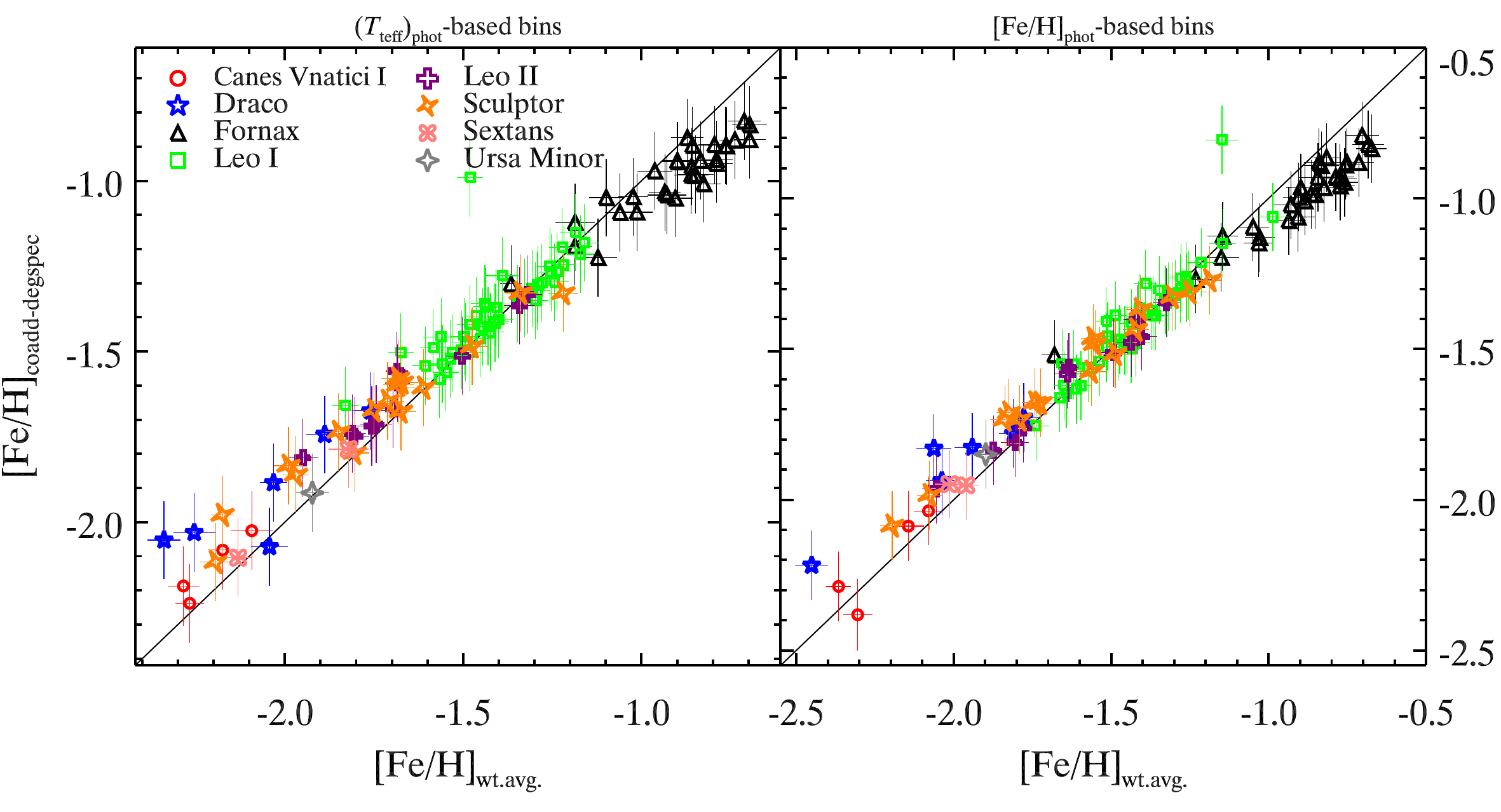}
	 \caption{
     The comparison between weighted-average
    metallicity and metallicity from co-added degraded spectra for 8 dSphs in the Milky
    Way. The symbols are same as in Figure~\ref{fig:fehcomp}. The value of the vertical axis represents co-added results and horizontal axis is weighted-average results. For weighted-average
    abundances, we measured every star's abundance and averaged the abundances, weighting by the elemental mask weights
    which are used in the combining of individual spectrum in co-adding spectrum.
    We selected stars with $\log \it g \rm \leq 1.4$ for all 8 dSphs stars and binned stars by
     photometric effective temperature (left) and photometric metallicity (right) respectively. Figure~\ref{fig:fehcomp}
    presents the same information based on non-degraded spectra with the observed S/N.
    \label{fig:degdsph}}
\end{figure}

\begin{figure}[t]
  \epsscale{1.2}

  \plotone{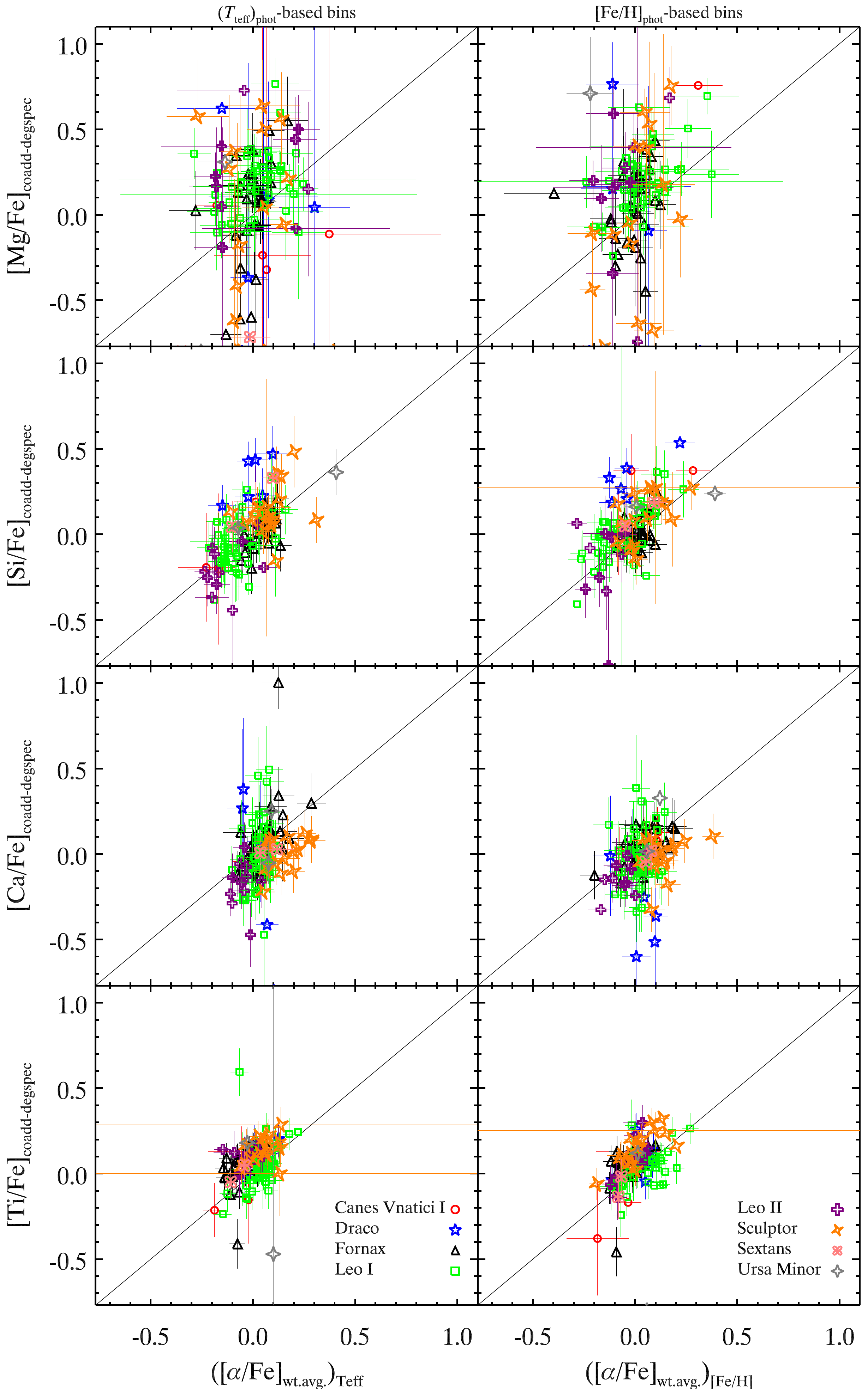}
	 \caption{Comparison for $\alpha$-element
    abundances between weighted-average abundances and co-added degraded
    spectral abundances for four $\alpha$-elements, Mg, Si, Ca, and Ti
    (top to bottom). Stars used for these plots are from the 8 dSph
    galaxies. Points in the left panels are stars binned by
    $T_{\rm eff}$ and right hand panels are binned by \feh$_{\rm phot}$. The
    symbols are same as in Figure~\ref{fig:degdsph}.  Figure~\ref{fig:alphacomp}
    presents the same information based on non-degraded spectra with the observed S/N.
    \label{fig:degdspha}}
\end{figure}

The aim for developing our co-add spectral technique is to use it for abundance measurements of medium-resolution spectra
of RGB stars in M31's dwarf satellite galaxies. The S/N of these
stars, however, are far lower (with a typical S/N of $<$ 10/pixel) than that for the
stars in this test. To complete this test, we reran our codes with
degraded spectra and compared these measurements with the weighted-average results derived in
\S4.1. We have randomly chosen 24 RGB stars of NGC185 and calculated
 the weighted-average S/N of these 24 spectra with Equation~\ref{eq:co-addsn}. 
We reference $\tt arm\_addnoise$ written by Marble (2004) to degrade the high S/N spectra. 
We compare the errors ($\sigma_{\rm MW}$) of individual reduced high S/N spectra with the rebinned errors 
($\sigma_{\rm M31}$) from the weighted-average S/N of 24 M31 spectra, to determine the pixels whose $\rm S/N_{\rm MW}$ are higher 
than $\rm S/N_{\rm M31}$ to degrade. We degrade the high S/N spectrum by adding new noise to the spectrum as below:
\begin{equation}
f_{\rm new}=f_{\rm old}+n_{\rm Gaussian}\cdot(\sqrt{(\frac{S/N_{\rm M31}}{S/N_{\rm MW}})^{-2}-1}\cdot \sigma_{\rm MW}) 
\end{equation} 
\begin{equation}
\sigma_{\rm new}=\sigma_{\rm MW}\cdot \frac{S/N_{\rm MW}}{S/N_{\rm M31}}
\end{equation}
where $f_{\rm old}$ and $f_{\rm new}$ are the fluxes of high S/N spectrum before and after added noise, respectively.  
$n_{\rm Gaussian}$ is Gaussian noise. $\sigma_{new}$ is the new error for the new flux. We degraded all the spectra in 
each bin individually before we ran the co-addition. 
We remeasure all the chemical abundances described in \S3.4 and \S3.5 but with the degraded spectra.
Figure~\ref{fig:deggc} and Figure~\ref{fig:deggca} present the comparison of \feh\ and \afe\ for GCs, respectively.
For comparison purpose, we also plot the high-resolution spectroscopic metallicities of GCs in red. M5 gives the best match, and
NGC2419 shows a discrepancy in which the coadd metallcity is less than
that determined from HRS by around 0.3 dex. \afe\ shows larger
scatter compared with the \afe\ measured from spectra without degrading, especially for \mgfe.
The comparisons for dSphs are shown in Figure~\ref{fig:degdsph} and Figure~\ref{fig:degdspha}.
The bias trends are larger than the trend in Figure~\ref{fig:fehcomp} in which the co-added degraded spectra tend to give metal-poorer
results for the metal-rich weighted-average groups but the inverse is
true for the metal-poor weighted-average groups. The matches
are good, however, for both binning approaches. The \afe\ measured
from degraded spectra with low S/N like in M31 satellite RGB stars
do reflect the mean value of these bins of stars. \tife\ shows a good match while the limited neutral lines
for Mg with such low S/N make the accuracy worse.
%%%%%%%%%%%%%%%%%%degspec GCs

%%%%%%%%%%%%degspec dSphs

We acknowledge that the procedure we are using to degrade the MW GC and dSph
spectra does not exactly mimic the situation with the coaddition of the low
S/N M31 spectra. The MW spectra were degraded after sky subtraction and
continuum division, whereas these steps may be more difficult for the M31
spectra given their low S/N.

\subsection{Discussion}

The discrepancy between the co-added and weighted average abundances indicates
 the intrinsic degeneracy in stellar parameter space,
 and it is hard for us to disentangle this degeneracy just using our current data sets.
Uncertainties in the measurements also can cause discrepancy.
Understanding causes of the bias is essential for future abundance analysis once
 we apply the co-addition technique to RGB stars in the Andromeda system,
 for this discrepancy will surely be present in those abundance
 measurements. We also rate the two binning scenarios in this part.

\subsubsection{[Fe/H]}

Both GCs and dSphs display trends in which measured metallicities from
co-added spectra are likely to be higher than the
weighted-average metallicities, except for NGC7492 and Fornax.
Globular clusters are simple stellar systems with old age and a single metallicity.
In our GCs sample, M15, M79, and NGC2419 have more than one bin and these bins show
a scatter covering an expanded range in
$\rm \langle S/N _{\rm coadd}\rangle$, $\langle \it T_{\rm eff} \rangle$,
and $\langle \log \it g \rangle$ in two binning approaches.
The $\langle \it T_{\rm eff} \rangle$ shows more impact on the scatter, which
increasing $\it T_{\rm eff}$ result bigger scatter. The rest of
GCs with one bin show a good match between coadd and weighted-average results.
The S/Ns for GCs are too high to assess its impact in Figure~\ref{fig:sngc}, but for the degraded spectra
of GCs, Figure~\ref{fig:deggc} demonstrates that even for spectra having S/N as low as
10/pixel, the coaddition technique can still produce reliable measurements, and the differences between coadded
 metallicites and HRS metallicites are acceptable.
For the scatter of bins of M15, M79 and NGC2419, we surmise
the main reason is measurement uncertainty.

The dSphs have more complicated and extended
star formation histories with spreads in their metallicity distributions.
Like GCs, S/N does not seem to be a dominant factor in the measurement bias
for both high S/N coadd spectra in the top panels of Figure~\ref{fig:sn} and degraded coadd spectra
in Figure~\ref{fig:degdsph}. The mean effective temperature, as (top panels)
Figure~\ref{fig:teff} shows, is tightly related to the metallicity residuals,
especially for $\it T_{\rm eff}\ge \rm 4300$ . $\log \it g$ seems to
have a slight impact
on the residuals of stars binned by $\it T_{\rm eff}$ as Figure~\ref{fig:logg} shows, which
just demonstrates what we note in \S3.4, which is that our synthetic spectral measurements use neutral
 metal lines which are insensitive to surface gravity. The detailed comparisons of medium-resolution
results and HRS results for individual stars of dSphs can be found in Kirby et al.\ (2010).

Effective temperature has an important impact
on the trends of residuals.
Metallicities determined from the co-added spectra of
metal-rich stars with low effective temperature (e.g., Fornax) tend to be more
 metal-poor compared with metallicities derived from the weighted average method.
As the effective temperature increases, the metallicity
measurement residuals for co-added spectra become more metal-rich.
In spectral synthesis,
when $\it T_{\rm eff}$ increases, it affects \feh\ because
more metals are needed in the synthetic atmosphere to compensate for the weakening lines
caused by increasing temperature (KGS08), and vice versa for metal-rich stars with
lower effective temperature. In spectral co-addition, this effect
seems to be amplified, probably because
there is a range of effective temperature for each bin, and the
stars with higher effective temperature favor
the more metal-rich synthetic one to compensate the temperature-induced
weakening lines, and vice versa.
Schlaufman et al.\ (2011) noticed the effect of the effective temperature on the
co-addition method, and they employed 500K bins in their stellar co-additions.
Our bins of stars do not have a fixed range in
effective temperature but $\it T_{\rm eff}$ for each bin are
in close proximity, a range smaller than 500K, as shown in
the left panels of Figure~\ref{fig:binplots}.
 Since $\it T_{\rm eff}$ and $\log \it g$ are correlated, we expect
	$\log \it g$ would have the same impact on metallicity, but it shows a less prominent
impact than $\it T_{\rm eff}$ on metallicity. The intrinsic metallicity dispersion of
individual systems also plays an essential role in the individual
discrepancy of different stellar systems. Schlaufman et al.\ (2011)
determined the average metallicity of M13 and M15 by co-addition but
using metal-poor main sequence turnoff stars from the Milky Way halo.
At lower spectral resolution, they measured $\feh =-1.7 \pm 0.15$ for M13 and
$\feh =-2.4 \pm 0.2$ for M15. Our results are more accurate, and the bias is
less than 0.01 for M13 and less than 0.06 for M15 (Figure~\ref{fig:gcsfeh}).

\subsubsection{[$\alpha$/Fe]}

Compared to the iron abundance, determining individual $\alpha$-element
abundances is more difficult, especially for distant stars with low S/N.
For both high S/N spectra and degraded spectra,
the comparisons of \afe\ between co-added and weighted average results show
conspicuous scatter, especially for \mgfe. For
GCs, the most biased bins are from the metal-poor systems M15 and
NGC2419, as Figure~\ref{fig:gcsalpha} and Figure~\ref{fig:deggca} show,
indicating that the measurements of $\alpha$-elements,
particularly \mgfe, are very sensitive to metallicity.
Increasing $\langle \it T_{\rm eff} \rangle$ still has an impact on the scatter in
 multi-bin GCs. M13, with the lowest  $\langle \it T_{\rm eff} \rangle$, presents an
obvious bias in \cafe\ and \tife. Compared to \mgfe, however, these biases are still acceptable, even for
the results derived from degraded spectra. S/N plays a more important role in $\alpha$ measurements,
 particular for \mgfe\ and \tife\ (Figure~\ref{fig:sngc}). The same impact of S/N also shown in
Figure~\ref{fig:sn} for dSphs. Another main reason for the scatter is the measurement uncertainty, especially
for \mgfe\ which can be seen in the measurements of degraded spectra (Figure~\ref{fig:deggca}).
For dSphs, the number of bins makes the impacts of different parameters on the residual biases more clear.
\sife, \cafe\ and \tife\ show good agreement between co-addition and
weighted average abundances in Figure~\ref{fig:alphacomp}. We suspect
that the large scatter for Leo~I
results from the low S/N and relatively high and extended coverage in $\langle \it T_{\rm eff} \rangle$.
For the outliers, CVnI, Sculptor, and Sextans, uncertainty in photometrically estimating
$\it T_{\rm eff} $ and \feh, which we used to bin the stars, is another
possible biasing source (Figure~\ref{fig:fehcompphot}).
Comparing Figures~\ref{fig:sn}, \ref{fig:teff}, and \ref{fig:logg}, it seems
that S/N, $\it T_{\rm eff}$, and $\log \it g$ are all possible
 reasons for the residual biases, especially for stars binned by
$(\it T_{\rm eff})_{\rm phot}$. Digging deeper, $\log \it g$ is a
function of stellar mass and radius, but the mass and radius of
star are also related to $\it T_{\rm eff}$.
%Both of these parameters, combined with \feh, shape the spectrum,
%plus the impact of S/N.
If \feh\ is fixed, increasing $\it T_{\rm eff}$ will make absorption lines more shallow and
narrow, but increasing $\log \it g$ will broaden these lines. If we consider
variation in \feh, then it becomes even more
difficult to distinguish which matters most. These small features affect the
abundance measurements for both individual and co-added spectrum.
Although stars in each bin have similar properties, the exact values of these
important stellar parameters are different. We used
the same chemical abundances to select the synthetic spectra for each
bin, but the photometric $\it T_{\rm eff}$ and $\log \it g$ values used
were those of individual observed stars. Considering the sensitivity of
$\alpha$ elements to the metallicity, this
assumption may play a role in the large scatter of \afe.
Compared to other $\alpha$ elements, \mgfe\ is the most difficult one
to measure accurately
because of weak absorption lines.
Magnesium is a product of Type~II supernova, and it is the least visible of the
$\alpha$ elements in the DEIMOS spectra (KGS08). The elemental mask for
[Mg/Fe] contains a limited number of absorption lines compared with other elemental
masks, therefore results in a larger uncertainty for \mgfe.

All 8 dSphs in this work have significant intrinsic
spread in metallicity (Figure~\ref{fig:fehcomp}). The
intrinsic spread in the abundance distribution of
dwarf satellite galaxies indicates their extended star formation
history (Venn et al.\ 2004; Helmi et al.\ 2006; Cohen \& Huang 2009;
Kirby et al.\ 2009). The target selection of a stellar system which has radial
metallicity gradient like Sculptor (Tolstoy et al.\ 2004;
Walker et al.\ 2009; Kirby et al.\ 2009) may influence the measured
abundance distribution for the
co-added spectra method. Therefore, when we apply co-addition
to dSphs, the effect of a metallicity spread on the $\alpha$-elements
abundance measurements with limited absorption lines need to be considered
carefully. Particularly for the stars in M31 satellite galaxies with low S/N,
Figures~\ref{fig:deggca} and \ref{fig:degdspha} show similar scatter amplitudes,
demonstrating that we can estimate the mean values and trends for $\alpha$ elements from
low S/N spectra with the coaddition technique.

\subsubsection{[$\alpha$/Fe] versus [Fe/H]}

\begin{figure}
  \epsscale{1.2}
  \plotone{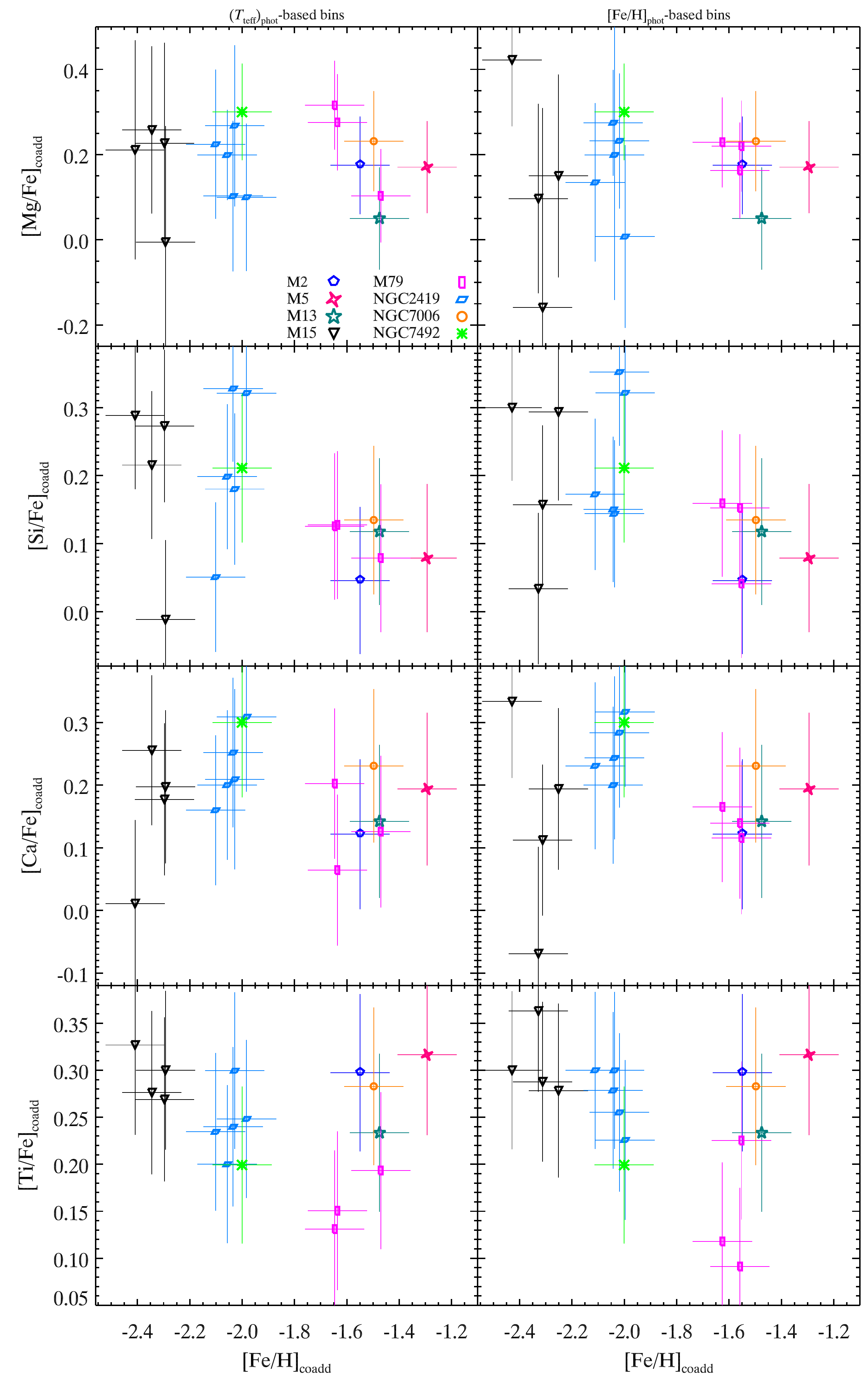}
  \caption{Element abundances for 8 GCs from co-added spectra. The
		left figure represents results binned by photometric effective
		temperature and the right figure is binned by
		photometric metallicity.\label{fig:afehgc}}
\end{figure}

\begin{figure}
  \epsscale{1.2}
  \plotone{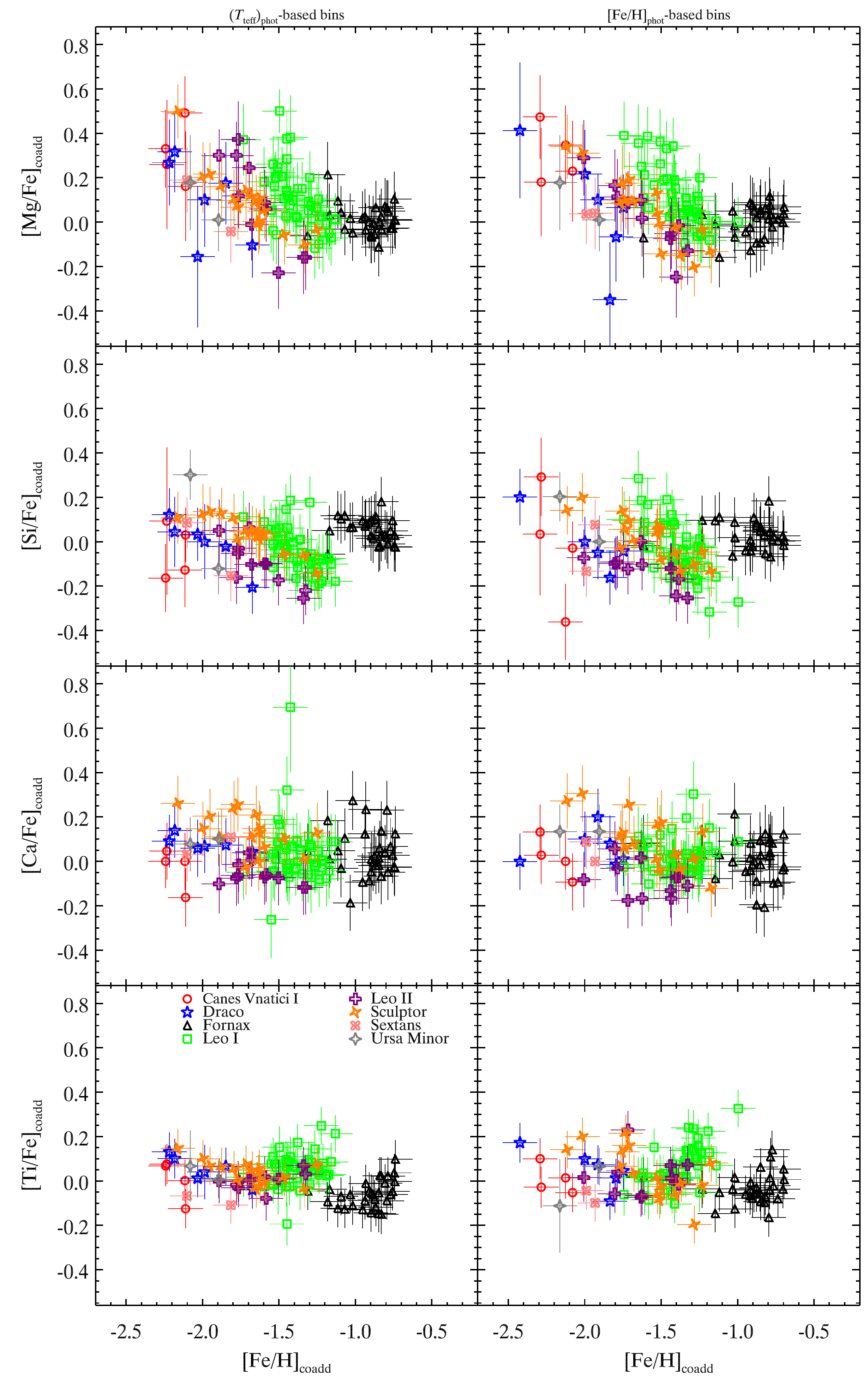}
  \caption{Multi-element abundances for 8 dSphs from co-added spectra. The
		left figure represents results binned by photometric
    effective temperature and the right figure is binned by photometric
    metallicity. The symbols are the same as in Figure~\ref{fig:fehcomp}.\label{fig:afeh}}
\end{figure}

\begin{figure}
  \epsscale{1.2}
  \plotone{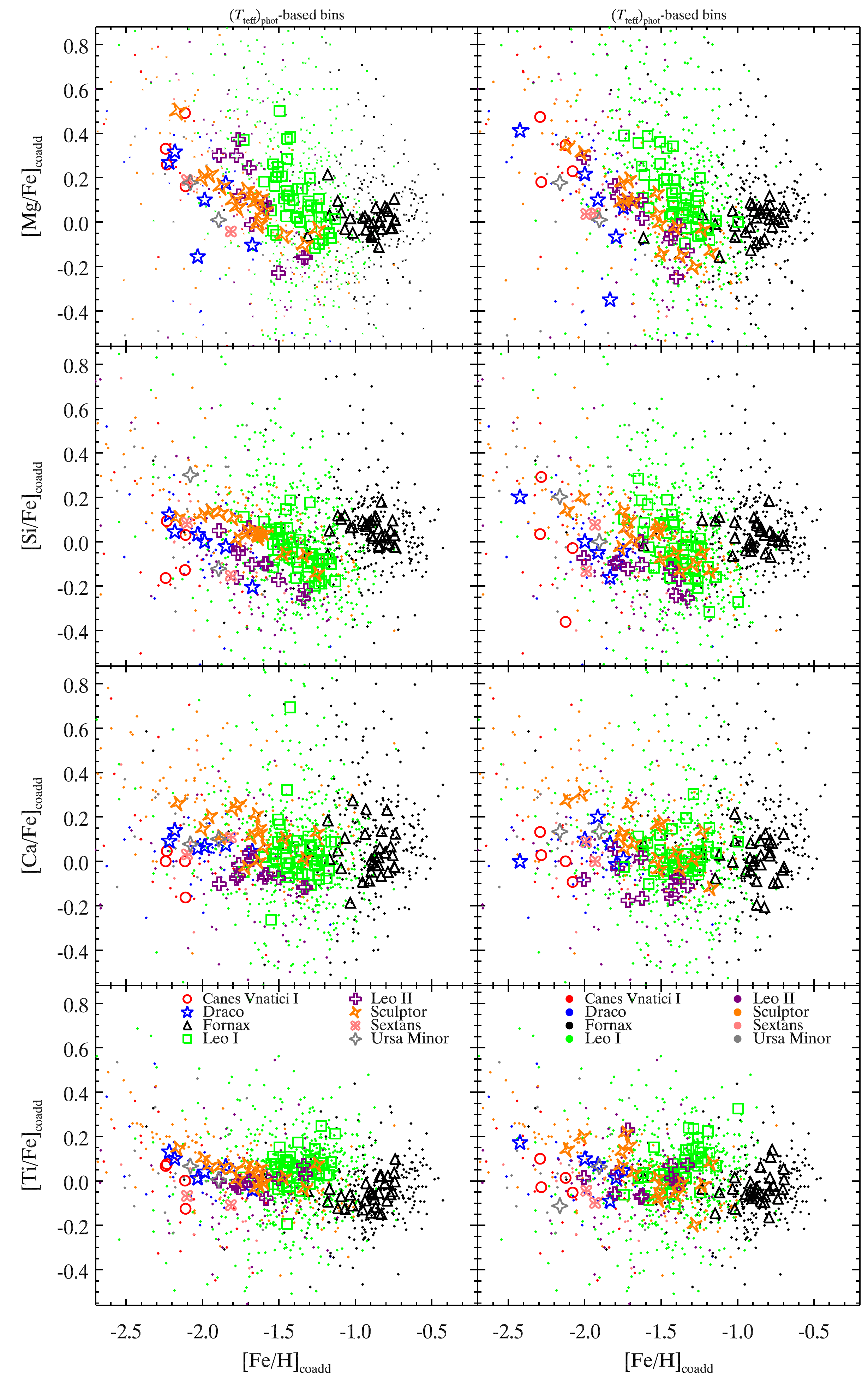}
  \caption{The same as in Figure~\ref{fig:afeh} but abundances of
		individual stars in each bin have been added in (the colorful
		points).\label{fig:afehstars}}
\end{figure}

Figure~\ref{fig:afehgc} and Figure~\ref{fig:afeh} present the individual $\alpha$-element
distribution versus metallicity determined from co-added
spectra for 8 GCs and 8 dSphs, respectively. The trends
are clearly shown for the majority of bins, and only Leo~I and Fornax display a very
blurred trend in \cafe\ distribution. We add individual stars'
abundances in the figures as Figure~\ref{fig:afehstars} shows. The distributions of \tife\ from Leo~I and
Fornax seem to display an upward tendency with increasing metallicity
compared with previous results shown by individual stars (Kirby et al.\ 2011b, their Figure~13).
We speculate that it is mainly because our surface
gravity restriction has removed fainter RGB stars, which are located
in the metal-rich region in Figure~\ref{fig:afeh}. Yet, we still
consider the co-added spectra method as
an efficient and feasible tool to proceed the detailed multi-element abundance
measurement with medium-resolution spectra. Based on the detailed $\alpha$-element
abundance analysis (Shetrone et al.\ 2001, 2003; Venn et al.\ 2004;
Kirby et al.\ 2011b), we expect to extend our understanding and insight into the
star formation history and galaxy evolution to the Andromeda galaxy
system and even farther systems with this technique.

\subsubsection{Binning Scenarios}

By comparing two binning algorithms, we can assess which one better
reflects the true values of groups of stars. The
two parameters we used to bin, $\it T_{\rm eff}$ and \feh$_{\rm
  phot}$, are primarily estimated from isochrone-fitting on
CMD. Stellar parameters estimated from photometry are far less
accurate than the ones derived by spectroscopy. On the other hand,
some assumptions must be made to constrain the parameter space in the CMD. Here
we have assumed all RGB stars have an age of 14 Gyr for both GCs and dSphs. For
the dwarf spheroidal galaxies, this assumption is too simple because
some dSphs are known to have extended star formation histories compared
with GCs (Harris 2001; Puzia 2003; Harris et al.\ 2006).
Sculptor and Sextans are dominated by old populations
(Orban et al.\ 2008), which should show observational properties
similar to Galactic globular clusters.
However, Shetrone et al.\ (2003) proved that Sculptor has a spread in age
of at least 4 Gyr. They also demonstrated that Fornax has a complicated
star formation history from 15 Gyr to 0.5 Gyr ago, Carina and Leo~I
have an age span of $4-7$ Gyr and  $2-7$ Gyr respectively.
de Boer et al.\ (2012) combined deep photometry with high resolution
spectroscopy to study the detailed star formation history and estimate
ages for individual red giant branch stars in the Sculptor. They found
that old ($>$10 Gyr) metal-poor stars, and younger, more metal-rich
populations are both present in Sculptor. By finding a knee in the
$\alpha$-element distribution at age around 10.9~Gyr they surmised that
SNe~Ia enrichment began ($2 \pm 1$ Gyr) after the beginning of star
formation in Sculptor. Furthermore, comparisons of the metallicities
of the Milky Way dwarf spheroidal galaxies derived from spectroscopy
versus photometry have indicated
a discrepancy between the two
techniques (Lianou et al.\ 2011). Lianou et al.\ (2011) demonstrated
that the assumption of old ages for RGB stars in dwarf satellite
galaxies leads to a bias toward metal-poor estimates if the ages of
the stars are lower, as is possibly the case for Fornax and Leo~II. The
ages of stars are always difficult to estimate, although one technique
is based on using the
spectroscopic metallicity, and then measuring the age using photometry
(KGS08).

In this work, the age affect is not enough
to highly bias our results, despite the age-spread in dSphs. In order to test the impact of
age, we have tried three ages in our photometric metallicity determination, 7 Gyr, 9 Gyr, and 14 Gyr,
and we found no significant difference in these ages. Then we choose 14 Gyr to set to all the test RGB stars
and make comparison of metallicity derived from co-added spectra and isochrone-fitting photometrically
for GCs and dSphs, as Figure~\ref{fig:fehcompphot} shows. There is a good reason, however, that $\feh_{\rm phot}$ is less than
$\feh_{\rm coadd}$ for Fornax.  The average age is significantly younger than 14 Gyr, which
means the RGB stars are bluer than the 14 Gyr isochrone. So $\feh_{\rm phot}$
will be too low when we assume the age is 14 Gyr. The uncertainty
in \feh$_{\rm phot}$ may affect the results if \feh$_{\rm phot}$ is significantly
different from \feh\ determined by spectroscopy. However, binning by
\feh$_{\rm phot}$ is still efficient, and Figure~\ref{fig:fehcomp}
shows that the span of
metallicities derived from stars binned by \feh$_{\rm phot}$ covers a larger
range than the one binned by $\it T_{\rm eff}$, meaning that we would
most likely discover extremely metal-poor stars by this way in more
distant stellar systems. These
intrinsic extremely metal-poor stars play a crucial role in chemical
evolution of these satellite galaxies and accretion history of galactic
stellar halo (Tolstoy et al. 2001; Helmi et al. 2006; Kirby et al.
2008b).

In the parameter space, the residuals of abundance binned by $\it
T_{\rm eff}$ span a larger range, as
shown in Figure~\ref{fig:teff} and Figure~\ref{fig:logg}, illustrating that these RGB stars
are very sensitive to $\it T_{\rm eff}$ and show a larger variance with
parameters such as S/N, $\it T_{\rm eff}$ and $\log \it g$. However,
the standard deviations of distributions show that chemical abundances
determined by \feh$_{\rm phot}$-binned stars are tighter and
similar to the weighted average abundances. The outliers in the
$\it T_{\rm eff}$-binned stars are expected to induce the main source
of the large scatter. In the future, we expect to combine these two
parameters together, and to explore
the impact of relationship between ages, photometric $\it T_{\rm eff}$
and $\log \it g$ and metallicity
on binning stars. Disentangling this problem will decrease the scatter
and improve the agreement between co-added results and weighted average
abundances.

\section{Summary and Applications}

In this paper we present a method for spectral coaddition of medium resolution spectra for
detailed measurements of multi-element
abundances. Our method overcomes the low S/N and
weak spectral features of distant RGB stars by co-adding spectra of
similar RGB stars. We impose a surface gravity restriction ($\log \it
g \rm \leq 1.40$) for RGB stars in GCs and dSphs
in order to enhance the quality of spectra. We group
stars in two ways, using photometric effective temperature estimates and
photometric metallicity estimates,
and then we determine detailed chemical abundances
(\feh, \mgfe, \sife, \cafe, \tife) with elemental masks on the whole
spectrum. We use more than 1300 well studied individual RGB stars from 8 GCs and 8 dSphs in the Milky Way to test the
feasibility and accuracy of this method. This work can briefly be
summarized as follows.

First, we determine effective temperature, surface gravity and
iron abundance by an isochrone-fitting technique on the CMD for
individual RGB stars in our sample, then
set these parameters as initial values for the following selection of
synthetic spectra for each observed one and the chemical abundance comparisons. Then, individual
chemical abundances for all RGB stars have been determined and compare
well with previous results (KGS08; Kirby et al.\ 2009, 2010).

Second, we make a cut ($\log \it g \rm \leq 1.4$) on candidate RGB stars so that our resulting sample is a good
 match to the M31 red giant stars to which we aim to apply this spectral co-addition method. Then the candidate stars
were binned using two photometric parameters,
effective temperature and metallicity. The science spectra belonging to one
bin were added together, weighted by the inverse variance within
specific elemental masks on the pixel level. We also selected a synthetic
spectrum for each observed spectrum in each bin based on its photometric effective temperature and
photometric surface gravity. The selected synthetic spectra have the same
metallicity and $\alpha$-element abundances for each bin. Then, we added these
synthetic candidate spectra together with the
same weights as we used for the observed ones.

Third, we used the Levenberg-Marquardt algorithm to select the best
fitting co-added synthetic spectrum for the co-added observed one by
minimizing $\chi^2$ between synthetic one and observed one for whole
spectrum on a pixel-to-pixel level in several iteratively steps. For
each bin we determine each elemental abundance, \feh, \mgfe,
\sife, \cafe, \tife, separately. Only one element was considered in
each run excluding bias probably induced by unavailable element
abundances of individual stars.

Fourth, we combined individual abundances in each bin weighted almost
the same way as we performed to the co-added spectra, to get the weighted average
abundances, and then we carried out
comparisons with co-added abundances. The co-added abundances agree
reasonably well with the expected values from the
weighted average abundances.

Fifth, for $(\it T_{\rm eff})_{\rm phot}$ versus $\feh_{\rm phot}- \rm
based$ binning scenarios, the abundance comparison and difference plots look similar for the
two scenarios. This means the abundance errors are dominated by factors
other than the spread of absorption line strength within a bin. For
metal poor stars the RGB isochrones are really vertical so the two
binning scenarios result in similar groups of stars. We prefer the
$\feh_{\rm phot}$ based binning scenario since our co-addition scenario
for synthetic spectra can account for a spread in $\rm T_{\rm teff}$ within a bin,
but not for a spread in \feh\ or \afe.

Last, the precision of the abundances measured from co-added MRS does not
appear to be a strong function of S/N over the range we have explored:
co-added spectral $\rm S/N \sim 200/pixel - 2000/pixel$ for GCs (see
Figure~\ref{fig:sngc}) and $\rm S/N \sim 50/pixel - 300/pixel$ for dSphs
(see Figure~\ref{fig:sn}). The S/N of the individual M31 RGB spectra to
which we plan to apply this co-addition method is typically much lower
than that of the individual spectra analyzed in this paper, but many
more stars are co-added together in the M31 bins so that the S/N of the
co-added M31 RGB spectra are expected to be comparable to that of the
co-added spectra used here (Kirby et al. 2010). The comparisons of metallicity
 and $\alpha$ elements derived from the degraded spectra with S/N comparable
to RGB stars of M31 satellite galaxies and weighted average results demonstrate
the feasibility of this technique.

We conclude that we can safely apply
this method of spectral coaddition to analyze chemical abundance patterns in M31
satellite galaxies using medium resolution spectra of RGB stars. This
will be a provide a useful start for detailed chemical abundance exploration beyond
the Milky Way system.

\acknowledgments

L.~Y. and E.~W.~P. gratefully acknowledge partial support from the Peking
University Hundred Talent Fund (985) and grants 10873001 and 11173003 from
the National Natural Science Foundation of China (NSFC). L.~Y. also
acknowledges support from the LAMOST-PLUS collaboration, a partnership
funded by NSF grant AST-09-37523, and NSFC grants 10973015 and
11061120454.

Support for this work was also provided by NASA through Hubble Fellowship
grant 51256.01 awarded to E.N.K. by the Space Telescope Science
Institute, which is operated by the Association of Universities for
Research in Astronomy, Inc., for NASA, under contract NAS 5-26555.

P.~G. acknowledges support from NSF grant AST-10-10039. He would like to thank
the staff of the Kavli Institute for Astronomy and Astrophysics at
Peking University for their generous hospitality during his collaborative visits.

L.~C. was supported by UCSC's Science Internship Program (SIP).

{\it Facilities:} \facility{Keck(DEIMOS)}.

\clearpage
\clearpage

\end{document}